\documentclass[review]{elsarticle}

\usepackage{lineno,hyperref,bm, amsmath,amssymb,amsthm,amsfonts,graphics,graphicx,epsfig,rotating,caption,subcaption,url, float, color}
\modulolinenumbers[5]

\journal{Journal of \LaTeX\ Templates}

\newcommand{\para}{\bigskip\noindent}
\newcommand{\bbeta}{\bm{\beta}}

\bmdefine\mmu{\mu}

\newcommand{\Bbeta}{\boldsymbol{\beta}}
\newcommand{\Btheta}{\boldsymbol{\theta}}
\newcommand{\BX}{\boldsymbol{X}}
\newcommand{\BY}{\boldsymbol{Y}}
\newcommand{\By}{\boldsymbol{y}}
\newcommand{\BQ}{\boldsymbol{Q}}
\newcommand{\BR}{\boldsymbol{R}}
\newcommand{\BU}{\boldsymbol{U}}
\newcommand{\BV}{\boldsymbol{V}}
\newcommand{\BI}{\boldsymbol{I}}
\newcommand{\BH}{\boldsymbol{H}}
\newcommand{\BA}{\boldsymbol{A}}
\newcommand{\BB}{\boldsymbol{B}}
\newcommand{\BC}{\boldsymbol{C}}
\newcommand{\BD}{\boldsymbol{D}}
\newcommand{\BSigma}{\boldsymbol{\Sigma}}
\newcommand{\I}{\mathbf{I}}
\renewcommand{\L}{\mathcal{L}}

\newcommand{\BLambda}{\boldsymbol{\Lambda}}
\newcommand{\ML}{\text{ML}}

\newcommand{\Beps}{\boldsymbol{\epsilon}}
\newcommand{\iid}{\; {\buildrel\rm iid\over \sim}\; }
\newcommand{\tTr}{\text{tr}}

\DeclareMathOperator*{\argmin.}{arg\,min}
\DeclareMathOperator*{\argmax.}{arg\,max}

\graphicspath{{C:/Users/jurre/Documents/StatisticalScience/Year2/MasterThesis/Artikel/Plots/}}
\begin{document}

\begin{frontmatter}

\title{Estimation of variance components, heritability and the ridge penalty in high-dimensional generalized linear models
\tnoteref{mytitlenote}}
\tnotetext[mytitlenote]{Supplementary Information available on ...}

\author[mymainaddress,mysecondaryaddress]{Jurre R. Veerman}

\author[mythirdaddress]{Gwena\"el G.R. Leday}

\author[mymainaddress,mythirdaddress]{Mark A. van de Wiel\corref{mycorrespondingauthor}}
\cortext[mycorrespondingauthor]{Corresponding author}
\ead{mark.vdwiel@vumc.nl}


\address[mymainaddress]{Dep. of Epidemiology \& Biostatistics, Amsterdam Public Health research institute, Amsterdam University medical centers, Amsterdam, The Netherlands}
\address[mysecondaryaddress]{Mathematical Institute, Leiden University, Leiden, The Netherlands}
\address[mythirdaddress]{MRC Biostatistics Unit, Cambridge University, Cambridge, UK}
\begin{abstract}
For high-dimensional linear regression models, we review and compare several estimators of variances $\tau^2$ and $\sigma^2$ of the random slopes and errors, respectively. These variances
relate directly to ridge regression penalty $\lambda$ and heritability index $h^2$, often used in genetics. Direct and indirect estimators of these, either based on cross-validation  (CV) or maximum marginal likelihood (MML), are also discussed. The comparisons include several cases of covariate matrix $\BX_{n \times p}$, with $p \gg n$, such as multi-collinear covariates and data-derived ones. In addition, we study robustness against departures from the model such as sparse instead of dense effects and non-Gaussian errors.

An example on weight gain data with genomic covariates confirms the good performance of MML compared to CV.
Several extensions are presented. First, to the high-dimensional linear mixed effects model, with REML as an alternative to MML. Second, to the conjugate Bayesian setting, which proves to be a good alternative. Third, and most prominently, to generalized linear models for which we derive a computationally efficient MML estimator by re-writing the marginal likelihood as an $n$-dimensional integral.
For Poisson and Binomial ridge regression, we demonstrate the superior accuracy of the resulting MML estimator of $\lambda$ as compared to CV. Software is provided to enable reproduction of all results presented here.

\end{abstract}


\end{frontmatter}


\date{}
\section{Introduction}
Estimation of hyper-parameters is an essential part of fitting high-dimensional Gaussian random effect regression models, also known as ridge regression.
These models are widely applied in genomics and genetics applications, where often the number of variables $p$ is much larger than the number of samples $n$, i.e. $p \ggg n$.

We initially focus on the linear model. The goal is to estimate error variance $\sigma^2$ and random effects variance $\tau^2$ or functions thereof, in particular the ridge penalty parameter, $\lambda = \frac{\sigma^2}{\tau^2}$, or
heritability index, $h^2 = \frac{p\tau^2}{p\tau^2 + \sigma^2}$.
Here, the ridge penalty is used in classical ridge regression to shrink the regression coefficients to zero \citep{Hoerl1970}, whereas heritability measures the fraction of variation between individuals within a population that is due their genotypes \cite{Visscher}. The estimators of $\sigma^2$ and $\tau^2$ can be used to estimate $\lambda$ or $h^2$, but also for statistical testing \cite{kang2008efficient}. We review several estimators, based on maximum marginal likelihood (MML), moment equations, (generalized) cross-validation, dimension reduction, or degrees-of-freedom adjustment. Some of these estimators are classical, while others have recently been introduced.

We systematically review and compare the estimators in a broad variety of high-dimensional settings. For estimation of $\lambda$ in \emph{low-dimensional} settings, we refer to \cite{Muniz2009, maansson2011poisson, kibria2016some}.
We address the effect of multi-collinearity and robustness against model misspecifications, such as sparsity and non-Gaussian errors. The comparisons are extended to the linear mixed effects model, with $q \ll n$ fixed effects added to the model and to Bayesian linear regression. The linear model part is concluded by a genomics data application to weight gain prediction after kidney transplantation.

The observed good performance of MML in the linear model setting was a stimulus to consider MML for high-dimensional generalized linear models (GLM).
MML is more involved here than in the linear model, because of the non-conjugacy of the likelihood and prior. Therefore, approximations are required, such as Laplace ones. While these have been addressed by others \cite{Heisterkamp1999, wood2011fast}, we derive an estimator which is computationally efficient for $p \ggg n$ settings. For Poisson and Binomial ridge regression, we demonstrate the superior accuracy of MML estimation of $\lambda$ as compared to cross-validation.

Our software enables reproduction of all results. In addition, it allows comparisons for one's own high-dimensional data matrix by simulating the response conditional on this matrix, as we do for
two cancer genomics examples. Computational shortcuts and considerations are discussed throughout the paper, and detailed at the end, including computing times.

\subsection{The Model}
We initially focus on high-dimensional linear regression with random effects. Variables are denoted by $j = 1, \ldots, p$ and samples by $i = 1, \ldots, n$. Then:
\begin{align}
\By_{n \times 1} = \BX_{n \times p}\Bbeta_{p \times 1} + \boldsymbol{\epsilon}_{n \times 1}\nonumber \\
\Bbeta_{p \times 1} \sim \mathcal{N}(0,  \tau^{2} \BI_{p}) \label{model} \\
\Beps_{n \times 1} \sim \mathcal{N}(0, \sigma^{2} \BI_{n}). \nonumber
\end{align}
Here, $\By = (y_{1},  \ldots,y_{n})$ is the vector of responses, $\Bbeta = (\beta_{1}, \ldots,\beta_{p})^{T}$ corresponds to the random effects and $\boldsymbol{\epsilon} = (\epsilon_{1}, \ldots,\epsilon_{n} )^T$ is a vector of Gaussian errors. Furthermore, $\BX$ is a fixed $n \times p$ matrix: $(\BX_1 \cdots \BX_n)^T,$ with $\BX_i = (x_{i1}, \ldots, x_{ip})^T$.

%
\subsection{Estimation Methods}
We distinguish three categories of estimation methods:
\begin{enumerate}
\item Estimation of functions of $(\sigma^2, \tau^2)$, in particular $\lambda= \frac{\sigma^2}{\tau^2}$ \cite[]{GolubGCV}, used in ridge regression to minimize $||\By - \BX\Bbeta||_2^2 + \lambda||\bbeta||_2^2$,  and heritability $h^2= \frac{p\tau^2}{p\tau^2 + \sigma^2}$ \cite[]{BonnetHiLMM}.
\item Separate estimation of $\sigma^2$ \cite[]{Cule2011, Cule2012}, possibly followed by plug-in estimation of $\tau^2$.
\item Joint estimation: estimate $\sigma^2$ and $\tau^2$ jointly
\end{enumerate}
Below, we discuss several methods for each of these categories. They have several matrices and matrix computations in common, which we therefore introduce first.

\subsection{Notation and  matrix computations}
Throughout the paper, we will use the following notation:
\begin{equation}
\begin{split}
\hat{\Bbeta} &= \hat{\Bbeta}_\lambda = \BC_\lambda\By = (\BX^T\BX + \lambda I_{p \times p})^{-1} \BX^T\By\ \ \text{i.e. the linear ridge estimator}\\
\BH &= \BH_\lambda = \BX\BC_\lambda = \BX(\BX^T\BX + \lambda I_{p \times p})^{-1} \BX^T\ \ \text{i.e. the hat matrix.}
\end{split}\label{not}
\end{equation}
Many of the estimators below require calculations on potentially very large matrices. The following two well-known equalities can highly alleviate the computational burden.

First, $\BC= \BC_{\lambda}$, and hence also $\hat{\Bbeta}$ and $\BH$, can be efficiently computed by using singular value decomposition (SVD).
Decompose $\BX = \BU_{n \times n} \BD_{n \times n} (\BV_{p \times n})^T$ by SVD, and denote $\BLambda_q= \lambda\BI_{q}$. Then,
\begin{equation}\label{svd}
\BC = (\BX^{T}\BX + \BLambda_p)^{-1}\BX^T = \BV(\BD^{2} + \BLambda_n)^{-1}\BD \BU^T.
\end{equation}

The latter requires inversion of an $n \times n$ matrix only.
Second, the following efficient trace computation for matrix products applies to $\tTr(\BH)=\tTr(\BX\BC_\lambda):$
\begin{equation}
\tTr(\BA_{p \times n}\BB_{n \times p}) = \sum_{i=1}^n\sum_{j=1}^{p} [\boldsymbol{A} \circ \boldsymbol{B}^{T}]_{ij}.
\label{trace}
\end{equation}


\section{Methods}
\subsection{Estimating functions of $\sigma^2$ and $\tau^2$}
\subsubsection{Estimating $\lambda$ by $K$-fold CV}\label{looCV}
A benchmark method that is used extensively to estimate $\lambda=\sigma^2/\tau^2$ is cross-validation.
Here, we use $K$-fold CV, as implemented in the popular \texttt{R}-package \texttt{glmnet} \cite{Friedman2010}. Let $f(i)$ denote the set of samples
left out for testing at the same fold as sample $i$. Then, CV-based estimation of $\lambda$ pertains to minimizing the
cross-validated prediction error:
\begin{equation} \label{CV}
\lambda_{cv} = \argmin._{\lambda}\{\sum_{i = 1}^{n}(y_i - \BX_i \hat{\beta}_{\lambda}^{-f(i)})^2\},
\end{equation}
where $\hat{\bbeta}_{\lambda}^{-f(i)}$ denotes the estimate of $\bbeta$ based on training samples $\{1,\ldots, n\} \setminus f(i)$ and penalty $\lambda$.
Note that for leave-one-out-cross-validation ($n$-fold CV) the analytical solution of (\ref{CV}) is the PRESS statistic \cite[]{AllenPRESS}.


\subsubsection{Estimating $\lambda$ by Generalized Cross Validation}
Generalized Cross Validation (GCV) is a rotation-invariant form of the PRESS statistic. It is more robust than the latter to (near-diagonal)
hat matrices $\BH_\lambda$ \cite[]{GolubGCV}.
For the linear model, the criterion is \citep{Hastie2008}:
\begin{equation} \label{GCV}
\text{GCV}(\lambda) = \sum_{i = 1}^{n}\bigg(\frac{y_{i} - \BX_{i}^{T}\hat{\boldsymbol{\beta}}_\lambda}{n-\tTr(\BH_\lambda))}\bigg)^2,
\end{equation}
where the trace of $\BH_\lambda$ can be computed efficiently by (\ref{trace}). Then, $\lambda_{\text{gcv}} = \argmin._{\lambda} \text{GCV}(\lambda)$.

\subsubsection{Estimating heritability by HiLMM} \label{H2}
Heritability is defined by $h^2=\frac{p\tau^2}{p\tau^2 + \sigma^2}$. A recent method which estimates heritability directly using maximum likelihood is proposed in \cite{BonnetHiLMM}.
Analogously to equation (\ref{MVN}), it is based on writing:
\begin{equation}\label{hilmm}
\By \sim \mathcal{N}(\mathbf{0}, h^2\sigma^{*2}\boldsymbol{R}+ (1 - h^2)\sigma^{*2} \BI_{n}),
\end{equation}
where $\sigma^{*2} = p\tau^2 + \sigma^2$ and $\boldsymbol{R} = \BX \BX^{T}/p$. Now, apply an eigen-decomposition to $\BR$:
$\BR = \BQ \boldsymbol{L} \BQ^T$.
Then, heritability is estimated by \cite{BonnetHiLMM}:
\begin{equation}\label{h2}
h^2 = \argmax._{h^2} \bigg(- \text{log}\bigg(\frac{1}{n}\sum_{i = 1}^{n}\frac{\tilde{y}_{i}^{2}}{h^2(\ell_i - 1) + 1}\bigg) - \frac{1}{n}\sum_{i = 1}^{n}(\text{log}(h^2(\ell_i-1)+1)\bigg),
\end{equation}
with $\ell_i$ and $\tilde{y}_{i}$ the $i$th element of $\boldsymbol{L}$ and $\tilde{\By} = \BQ^{T}\By$, respectively.
The authors provide rigorous consistency results for their estimator, as well as theoretical confidence bounds, also for mixed models and sparse settings.

\subsection{Estimation of $\sigma^2$}
The two methods below rely on an estimate $\hat{\Bbeta} = \hat{\Bbeta}_\lambda$, where $\lambda = \sigma^2/\tau^2$ is estimated by (G)CV. Then
 $\sigma^2$ is estimated conditional on $\hat{\Bbeta}.$ If desired, $\tau^2$ may then be estimated by $\hat{\tau}^2 = \hat{\sigma}^2/\hat{\lambda}.$
\subsubsection{Basic estimate}
A basic estimate of $\sigma^2$, and often used in practice, is given by \cite[]{Hastie1990}:
\begin{equation}
\hat{\sigma}^{2} = \frac{(\By-\BX\hat{\Bbeta})^{T}(\By-\BX\hat{\Bbeta})}{\nu}, \label{MSE}
\end{equation}
which is the residual mean square error. Here, the residual effective degrees of freedom \cite[]{Hastie1990} equals $\nu = n - \text{tr}(2\boldsymbol{H} - \boldsymbol{HH}^{T})$, with $\BH$ as in (\ref{not}).
We also considered \eqref{MSE} with $\nu = n - \text{tr}(\boldsymbol{H})$, as in \cite{hellton2018fridge}, which rendered similar, slightly inferior results.

\subsubsection{PCR-based estimate}
The estimator for $\sigma^{2}$ may also be based on Principal Component Regression (PCR). PCR is based on the eigen-decomposition $\BX^{T}\BX = \tilde{\BQ}\BD^2\tilde{\BQ}^T$. Denoting $\boldsymbol{Z} = \BX \tilde{\BQ}$ and $\boldsymbol{\alpha} = \tilde{\BQ}^T\Bbeta$, we have $\By = \boldsymbol{Z}\boldsymbol{\alpha} + \boldsymbol{\epsilon}$. Then, $\boldsymbol{Z}$ is reduced from $p$ columns to $r \leq \text{min}(n,p)$ principal components, a crucial step \cite[]{Cule2012}. Using the reduced model, $\sigma^{2}$ is estimated by the residual mean square error \cite[]{Cule2012}:
\begin{equation}\label{pcr}
\hat{\sigma}^{2}_{r} = \frac{(\By - \boldsymbol{Z}_{r}\hat{\boldsymbol{\alpha}}_{r})^{T}(\By - \boldsymbol{Z}_{r}\hat{\boldsymbol{\alpha}}_{r})}{n - r}.
\end{equation}

\subsection{Joint estimation of $\sigma^2$ and $\tau^2$}
\subsubsection{MML}
An Empirical Bayes estimate of $\sigma^2$ and $\tau^2$ is obtained by maximizing the marginal likelihood (MML), also referred to as model evidence in machine learning \cite[]{Murphy2012}. This corresponds to:
\begin{equation}\label{eq:MML}
\argmax._{\sigma^{2}, \tau^{2}} P(\By) = \argmax._{\sigma^{2}, \tau^{2}} \int_{\Bbeta} \mathcal{L}(\By; \Bbeta, \sigma^2)\pi(\Bbeta; \tau^2)d\Bbeta.
\end{equation}
Since $\By = \BX\Bbeta + \boldsymbol{\epsilon}$, $P(\By)$ is simply derived from the convolution of Gaussian random variables, implying $E[\By] = E[\BX\Bbeta] + E[\boldsymbol{\epsilon}] = \mathbf{0}$, and
$V[\By] = V[\BX\Bbeta] + V[\boldsymbol{\epsilon}]  = \BX \BX^{T} \tau^{2} + \sigma^{2} \BI_{n}$, so
\begin{align}
P(\By) &= \mathcal{N}(\By; \boldsymbol{\mu} = 0, \BSigma = \BX \BX^{T} \tau^{2} + \sigma^{2} \BI_{n}). \label{MVN}
\end{align}
This is easily maximized over $\sigma^{2}$ and $\tau^{2}$. Note that after computing $\BX \BX^{T}$ (\ref{MVN}) requires operations on
$n \times n$ matrices only.

\subsubsection{Method of Moments (MoM)}
An alternative to MML is to match the empirical second moments of $\By$ to their theoretical counterparts. From (\ref{MVN}) we observe that the covariances depend on $\tau^2$ only.
Hence, we obtain an estimator of $\tau^2$ by equating the sum of $y_iy_k$ to that of the theoretical covariances, $\BSigma_{ik} = \mathbb{E}[\By_i \By_k]$, with $\BSigma$ as in \eqref{MVN}. Then, with $\BSigma^{\BX} = \BX\BX^T$, an estimator for $\sigma^2$ is obtained by substituting $\hat{\tau}^2$ and
equating the sum of $y_i^2$ to the sum of theoretical variances, $\BSigma_{ii} =\mathbb{E}[\By_i^2]$:
\begin{equation}
\begin{split}
\hat{\tau}^2 &= \frac{\sum_{i \neq k}^{n,n} y_iy_k}{\sum_{i \neq k}^{n,n} \BSigma^{\BX}_{ik}}\\
\hat{\sigma}^2 &= n^{-1}\sum_{i=1}^{n}(y_i^2 - \hat{\tau}^2\BSigma^{\BX}_{ii})
\end{split}
\end{equation}
These equations also hold for non-Gaussian error terms, which could be an advantage over MML. Moreover, no optimization over $\sigma^2$ and $\tau^2$ is required, so MoM is computationally very attractive.

\section{Comparisons}
For the linear random effects model (ridge regression) we study the following settings:
\begin{itemize}
  \item $\Bbeta$ and $\Beps$ generated from model (\ref{model}), independent $\BX$
  \item $\Bbeta$ or $\Beps$ generated from non-Gaussian distributions, independent $\BX$
  \item $\Bbeta$ and $\Beps$ from model (\ref{model}), multicollinear $\BX$
  \item $\Bbeta$ and $\Beps$ from model (\ref{model}), data-based $\BX$.

\end{itemize}
As is common for real data, the variables, i.e. the rows of $\BX$, were always standardized for the $L_2$-penalty to have the same effect on all variables. All the results are based on 100 simulated data sets. Cross-validation is applied on 10 folds. Results from $n$-fold CV (leave-one-out) were generally fairly similar. We focus on the high-dimensional setting with $n = 100, p = 1000$,  with excursions to larger data sets and dimensions of real data. In all visualizations below the red dotted lines indicate true values. Moreover, values larger than 20 times the true value were truncated and slightly jittered. Discussion of all results is postponed to Section \ref{discres}.

\subsection{Independent $\BX$}
In correspondence to model \eqref{model} we sample :
\begin{equation}
\begin{aligned}
\label{basicmodel}
&\By_{n \times 1} = \BX_{n \times p}\Bbeta_{p \times 1} + \Beps_{n \times 1} &\qquad &\epsilon_i \iid \mathcal{N}(0, \sigma^{2})\\
&x_{ij} \iid \mathcal{N}(0, 1) &\qquad &\beta_j \iid \mathcal{N}(0,  \tau^{2}).
\end{aligned}
\end{equation}
Figures \ref{fig:indX}(a) and (b) display the results for $n=100, p=1000, \tau^2 = 0.01, \sigma^{2} = 10$ and for a large data setting $n=1000, p=15000, \tau^2 = 0.01, \sigma^{2} = 150$ (which both imply $h^2=0.5$).

\subsection{Departures from a normal effect size distribution}
We study the robustness of the methods against (sparse) non-Gaussian effect size distribution or error distribution.
In sparse settings, many variables do not have an effect. To mimic this, we simulated the $\beta$'s from a mixture distribution with a `spike' and a Gaussian `slab':
\begin{equation}\label{spikeslab}
\beta_j \iid p_{0}\delta_{0} + (1-p_{0})\mathcal{N}(0, \tau^{2}_{0}).
\end{equation}
Here, we set $p_{0} = 0.9, \tau^{2}_0 = 0.1,$ which implies $\tau^2 = \mathbb{V}(\beta_j) = \mathbb{E}(\beta_j^2) - \mathbb{E}(\beta_j)^2 = (1-p_0)\tau_0^2 = 0.01$, as in the Gaussian $\beta_j$ setting.
Moreover, we also considered:
$$\beta_j \iid \text{Laplace}(\mu = 0, b = 0.0707) \quad \text{and} \quad \beta_j \iid \text{Uniform}(a = -0.17, b = 0.17),$$
where again the parameters are chosen such that $E(\beta_j)=0$ and $\tau^2 = \mathbb{V}(\beta_j) = 0.01.$ Apart from $\bbeta$ all other quantities are simulated as in \eqref{basicmodel}.
Results are displayed for $\sigma^{2} = 10, \tau^{2} = 0.01, n = 100, p = 1000$ in Figure \ref{fig:indX}(c) for the Laplace (= lasso) effect size distribution and in Supplementary Figure 3 for the spike-and-slab and uniform effect size distribution.

\para
Moreover, we considered heavy-tailed errors by sampling
$$\epsilon'_{i} \iid \text{t}_4 \qquad \epsilon_{i}= (10/2)^{1/2}\epsilon'_{i},$$ where the scalar is chosen such that $\sigma^2=\mathbb{V}(\epsilon_i) = 10$, as in the Gaussian error setting. Apart from $\Beps$, all other quantities are simulated as in \eqref{basicmodel}.
Results are displayed in Supplementary Figure 3(c).

\subsection{Multicollinear $\BX$}
\subsubsection{Simulated $\BX$}
Next, the design matrix $\BX$ is sampled using block-wise correlation. We replace the sampling of $\BX$ in simulation model \eqref{basicmodel} by:
\begin{align}
\begin{split}
\label{multicoll}
\BX_{n \times p} &\sim \mathcal{N}(0, \boldsymbol{\Xi}),
\end{split}
\end{align}\label{multicol}
where $\boldsymbol{\Xi}$ is a unit variance covariance matrix with blocks of size $p^{*} \ll p$ with correlations $\rho$ on the off-diagonal.
Figure \ref{fig:Cor}(a) shows the results for $\rho = 0.5, p^{*} = 10, n = 100, p = 1000$.

\subsubsection{Real data $\BX$}
Finally, we consider the estimation of $\tau^2$ and $\sigma^2$ in a high- and medium-dimensional setting where $\BX$ are real data, with likely collinear columns.
The first data set (TCGA KIRC) concerns gene expression data of $p=18,391$ genes for $n=71$ kidney tumors. The second data set (TCPA OV) holds expression data of $p=224$ proteins for $n=408$ ovarian tumor samples.
Details on both data sets are supplied in the Supplementary Information. To generate response $\By$ we use model \eqref{basicmodel} with $\BX$ given by the data. Here, $\tau^2=0.01$ and $\sigma^2$
is set such that $h^2 = 0.5.$ Figures \ref{fig:Cor}(b) and (c) show the results.

\begin{figure}[H]
\caption{Results for independent $\BX$}
\centering
\begin{subfigure}{\textwidth}
  \centering
  \includegraphics[width=1\linewidth]{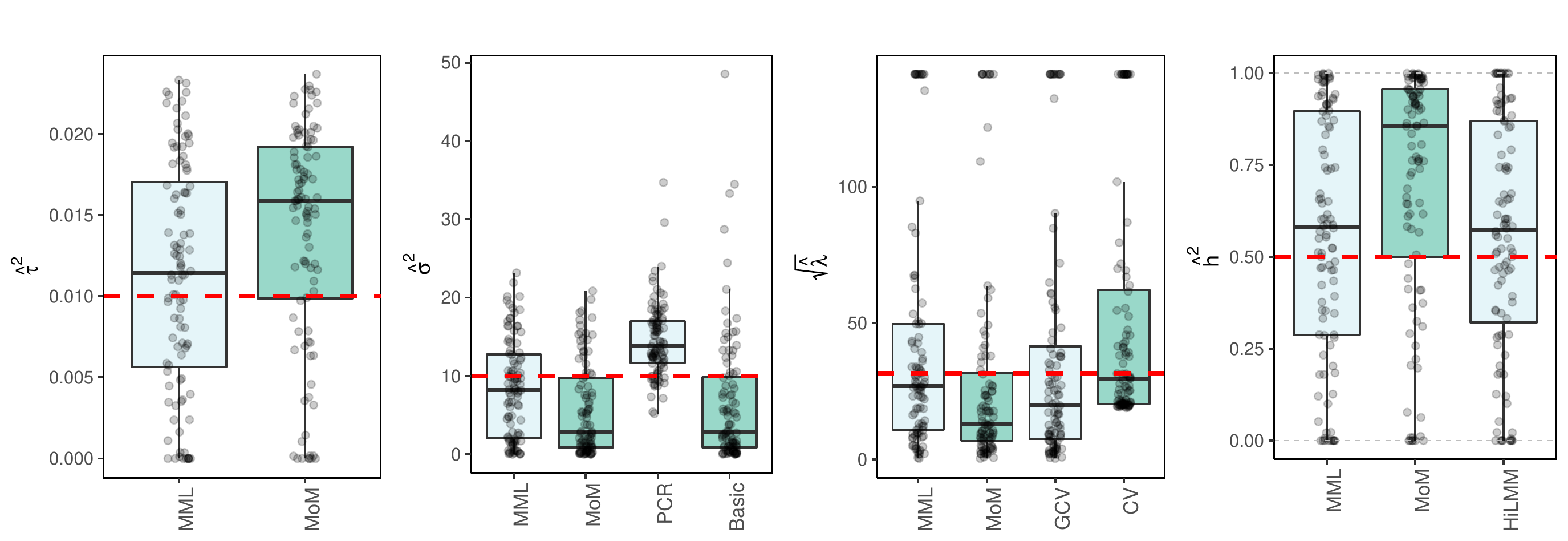}
  \caption{Standard setting: Gaussian $\beta$'s, $n=100, p=1000, \tau^2 = 0.01, \sigma^{2} = 10$}
\end{subfigure}%

\begin{subfigure}{\textwidth}
  \centering
  \includegraphics[width=1\linewidth]{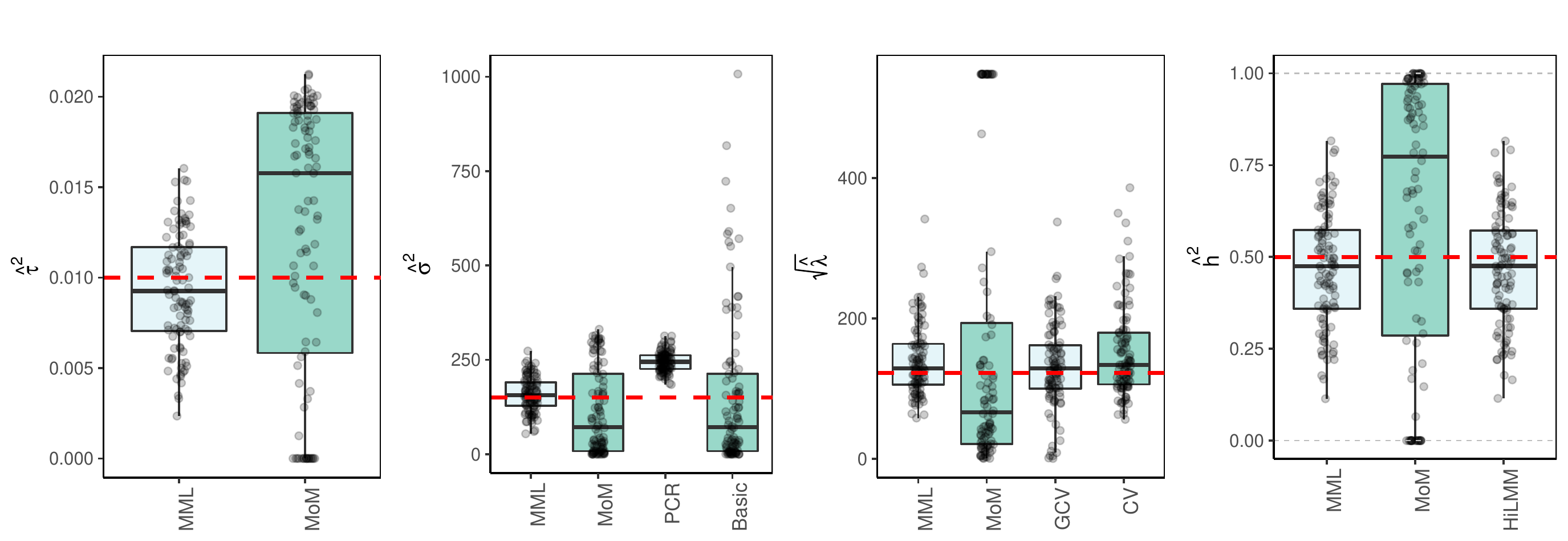}
  \caption{Large setting: Gaussian $\beta$'s, $n=1000, p=15000, \tau^2 = 0.01, \sigma^{2} = 150$}
\end{subfigure}

\begin{subfigure}{\textwidth}
  \centering
  \includegraphics[width=1\linewidth]{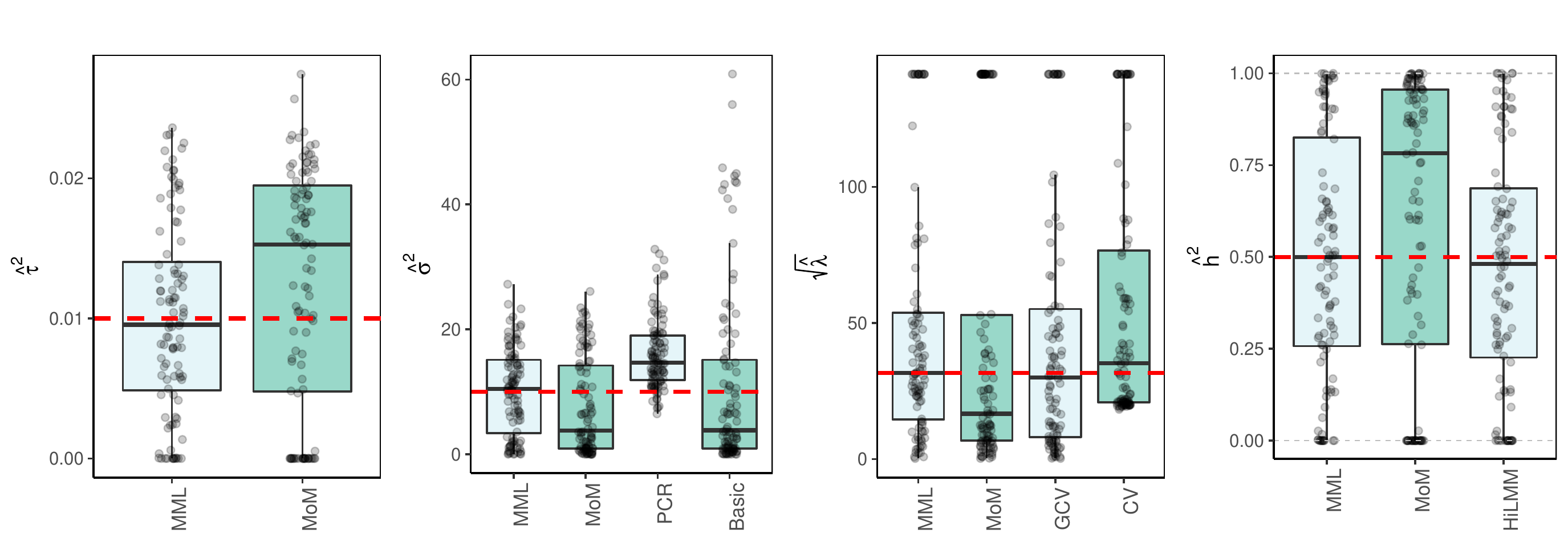}
  \caption{Lasso setting: Laplace $\beta$'s, $n=100, p=1000, \tau^2 = 0.01, \sigma^{2} = 10$}
\end{subfigure}
  \label{fig:indX}
\end{figure}



\begin{figure}[H]
\caption{Results for multi-collinear and real $\BX$}
\centering
\begin{subfigure}{\textwidth}
  \centering
  \includegraphics[width=1\linewidth]{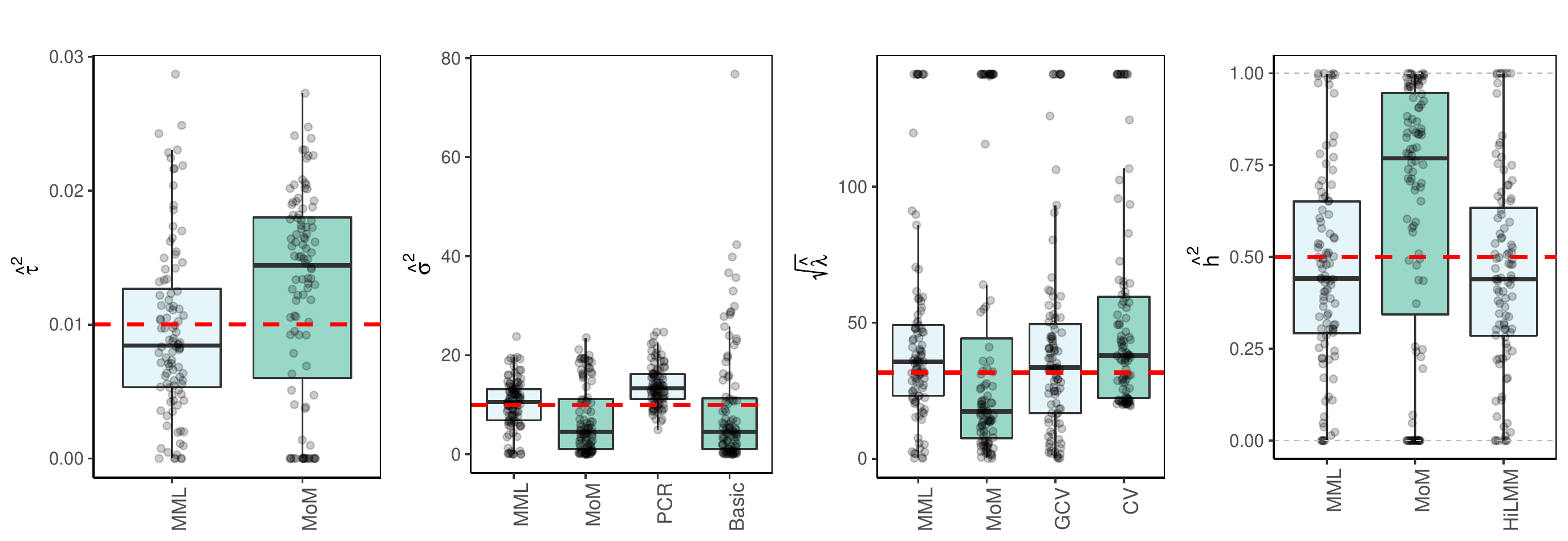}
  \caption{Multi-collinear $\BX$ setting: Gaussian $\beta$'s, $n=100, p=1000, \tau^2 = 0.01, \sigma^{2} = 10$}
\end{subfigure}%

\begin{subfigure}{\textwidth}
  \centering
  \includegraphics[width=1\linewidth]{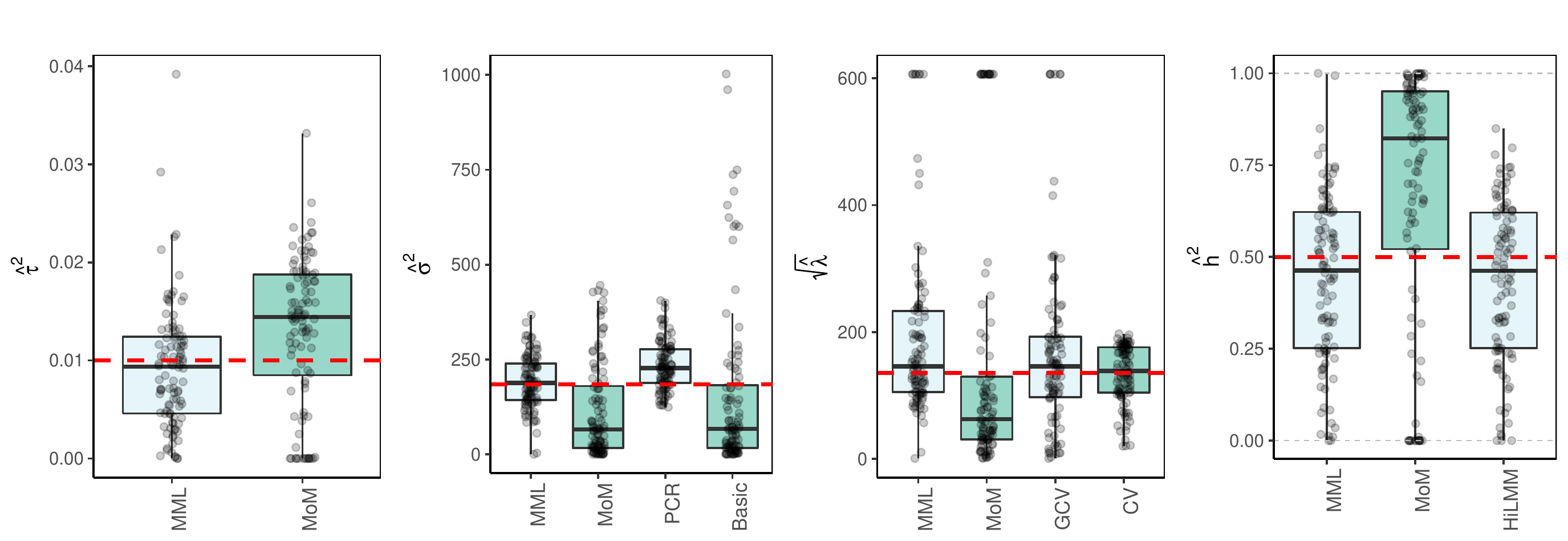}
  \caption{$\BX$ = TCGA KIRC data: Gaussian $\beta$'s, $n=71, p=18391, \tau^2 = 0.01, \sigma^{2} = 184$}
\end{subfigure}

\begin{subfigure}{\textwidth}
  \centering
  \includegraphics[width=1\linewidth]{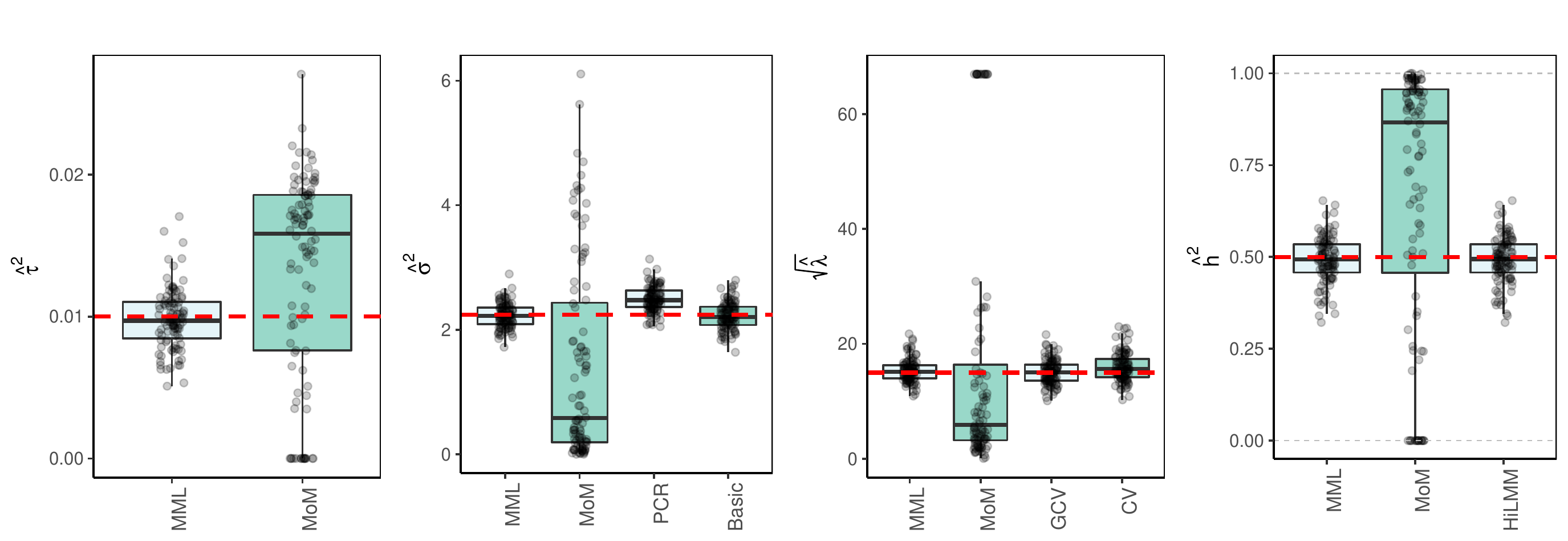}
  \caption{$\BX$ = TCPA OV data: Gaussian $\beta$'s, $n=408, p=224, \tau^2 = 0.01, \sigma^{2} = 2.24$}
\end{subfigure}
  \label{fig:Cor}
\end{figure}



\subsection{Discussion of results}\label{discres}
\subsubsection{MML vs MoM, Basic and PCR}
Figures \ref{fig:indX} and \ref{fig:Cor} and Supplementary Figure 3 clearly show superior performance of MML compared to MoM: both the bias and variability are much smaller for MML. Generally, MML also outperforms
the Basic and PCR estimators of $\sigma^2$. The PCR estimator approaches the performance of MML for the KIRC and TCPA data (Figures \ref{fig:Cor}(b) and \ref{fig:Cor}(c)), and the Basic estimator performs reasonably well
for the latter ($p<n$) data set. For other settings, the Basic estimator performs equally inferior as MoM. The results highlight the importance of joint estimation of $\sigma^2$ and $\tau^2$ in high-dimensional settings, because of their delicate interplay.

\subsubsection{MML vs GCV and CV}
For the estimation of $\lambda$ MML seems slightly superior to GCV and CV.  GCV shows more estimates that deviate towards too small values of $\lambda$ (e.g. Figures \ref{fig:indX}(b) and \ref{fig:Cor}(b), i.e. the large $p$ settings), whereas CV tends to render somewhat more skewed results, either to the right (Figures \ref{fig:indX}(a) and \ref{fig:indX}(c), \ref{fig:Cor}(a)), or to the left (Figure \ref{fig:Cor}(b)). For the spike-and-slab and uniform effects sizes and the $t_4$ errors the right-skewness of the CV-results is more pronounced (Supplementary Figure 3), indicating that minimization of the cross-validated prediction error (\ref{CV}) is more vulnerable to
non-Gaussian $\By$ than MML and GCV. Note that the Laplace setting (Figure \ref{fig:indX}(c)) relates directly to the lasso prior with scale parameter $1/\lambda_1$ \cite{tibshirani1996lasso}. The results indicate that MML with Gaussian prior could be useful to find the lasso penalty, or serve as a fast initial estimate by simply setting the lasso penalty $\lambda_1 = \sqrt(2)/\hat{\tau}$, which follows from the variance of the lasso prior.

\subsubsection{MML vs HiLMM}
For the estimation of heritability $h^2$ Figures \ref{fig:indX} and \ref{fig:Cor} and Supplementary Figure 3 show very comparable performance of MML and HiLMM. This similar performance is not surprising
given that both methods are likelihood-based. Hence, while reparametrizing the likelihood (\ref{hilmm}) is certainly useful to study it as function of $h^2$ \cite{BonnetHiLMM}, the reparametrization seems not beneficial for the purpose of estimating $h^2$. In addition, unlike HiLMM, MML also returns estimates of $\tau^2$ and $\sigma^2$.
Finally, comparing Figures \ref{fig:indX}(a) and \ref{fig:indX}(b) we observe that both MML and HiLMM clearly benefit from the larger $n$ and $p$.

\section{Data example}
We re-analyse the weight gain data, recently discussed in \cite{hellton2018fridge}. Details on the data are presented there, we provide a summary. The data consists of expression profiles of $n=26$ individuals with kidney transplants, where profiles consists of 28,869 genes as measured by Affymetrix Human Gene 1.0 ST arrays. The data is available in the EMBL-EBI ArrayExpress database (www.ebi.ac.uk/arrayexpress) under accession number
E-GEOD-33070.  It is known that kidney transplantation may lead to weight gain, and the study \cite{cashion2013expression} investigates whether gene expression can be used to predict this. Such a prediction can be used to
decide upon additional measures to prevent excessive weight gains. We reproduced the analysis by \cite{hellton2018fridge} as much as possible, including their prior selection of 1000 genes. Details on minor discrepancies, and an alternative analysis that accounts for the gene selection are discussed in the Supplementary Material. These did not affect the comparison qualitatively.

In \cite{hellton2018fridge}, the authors illustrate their focused ridge (\texttt{fridge}) method and compare it with conventional ridge.
In short, \texttt{fridge} estimates \emph{sample-specific} ridge penalties, based on minimizing a per sample mean squared error (MSE) criterion on the level of the linear predictor $\BX_i\Bbeta$.
Since $\Bbeta$ is not known, it is replaced by an initial ridge estimate, $\hat{\Bbeta}_{\lambda}.$ Their sample specific penalty then depends on $\BX_i$, and also on both $\hat{\lambda}$ and $\hat{\sigma}^2$.
The authors use GCV \eqref{GCV} to obtain $\lambda$, and a slight variation of \eqref{MSE} to estimate $\sigma^2$. They show that \texttt{fridge} improves upon GCV-based ridge estimation. We wish to investigate whether i) MML estimation of $\lambda = \sigma^2 /\tau^2$ also improves the performance of GCV-based ridge regression; and ii) whether MML estimation further boosts the performance of the \texttt{fridge} estimator. Here, predictive performance is measured by the mean squared prediction error (MSPE) using leave-one-out cross-validation (loocv).

\begin{figure}
\centering
\includegraphics[scale=0.7]{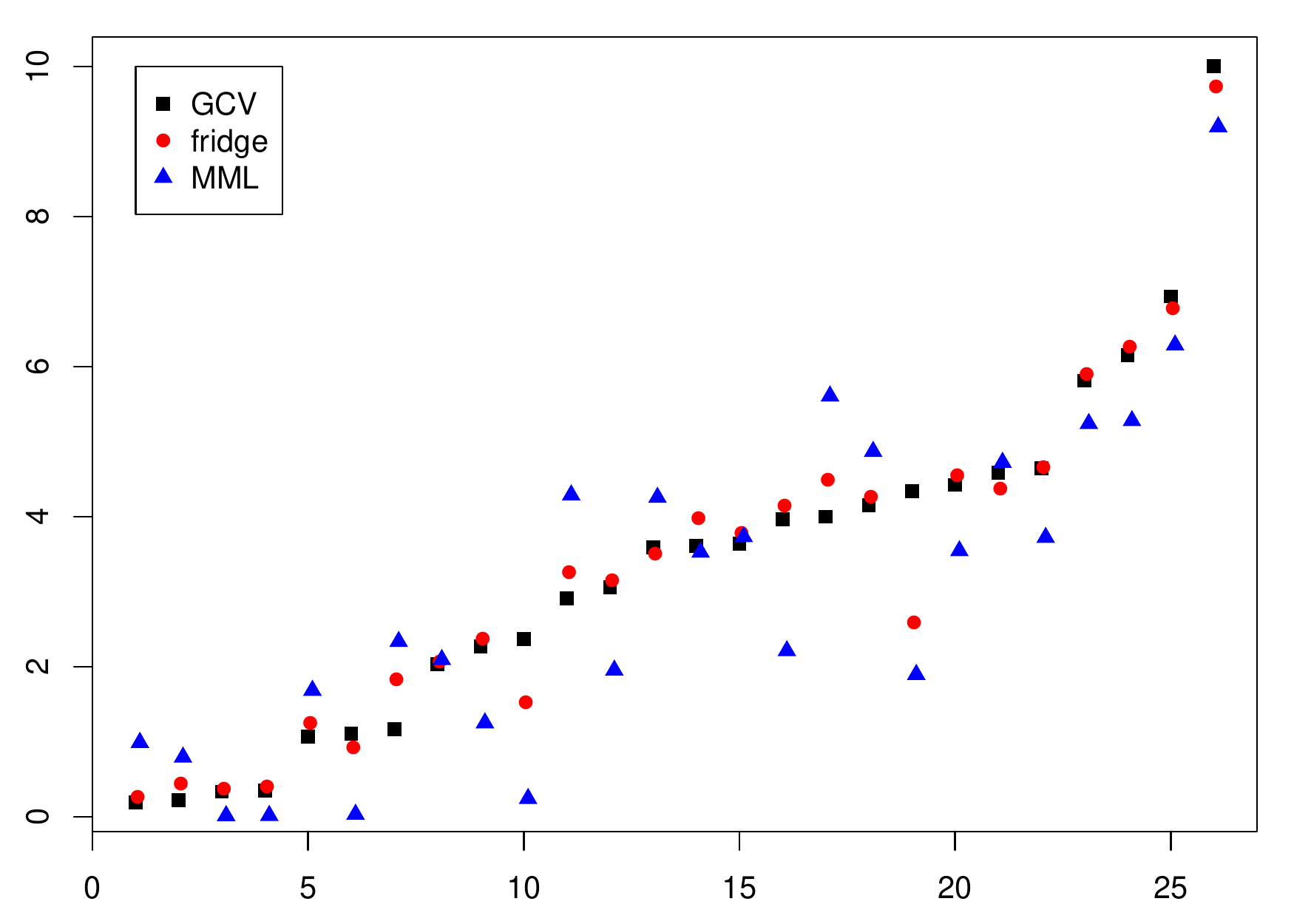}
\caption{Absolute prediction errors (obtained by loocv; y-axis) for ridge using $\lambda_{\text{GCV}}$, for \texttt{fridge} and for ridge using $\lambda_{\text{MML}}$. Sample indices (x-axis) are sorted by GCV results.}\label{weightgain}
\end{figure}

The estimates of MML differ markedly from those of GCV: $(\hat{\lambda}_{\text{MML}}, \hat{\sigma}^2_{\text{MML}}) = (0.77, 0.59)$, while $(\hat{\lambda}_{\text{GCV}}, \hat{\sigma}^2_{\text{GCV}}) = (20.92, 8.08)$.
Using $\hat{\lambda}_{\text{MML}}$ instead of $\hat{\lambda}_{\text{GCV}}$ for the estimation of $\Bbeta$ substantially reduced the mean squared prediction error: $\text{MSPE}_\text{MML} = 14.40,$ while
$\text{MSPE}_\text{GCV} = 16.38$, a relative decrease of 12.1\%. Using $\hat{\lambda}_{\text{GCV}}$, as in \cite{hellton2018fridge}, \texttt{fridge} also reduced the MSPE, but to a lesser extent:
$\text{MSPE}_\text{fridge} = 15.80,$ a relative decrease of 3.5\% with respect to $\text{MSPE}_\text{GCV}.$ Application of \texttt{fridge} using $\hat{\lambda}_{\text{MML}}$ did not further decrease
$\text{MSPE}_\text{MML}$, nor did it increase it. Possibly, the already fairly small value of $\hat{\lambda}_{\text{MML}}$ left little room for improvement. Figure \ref{weightgain} displays absolute prediction errors per sample and illustrates the improved prediction by ridge using $\lambda_{\text{MML}}$ (and to a lesser extent by \texttt{fridge}) with respect to ridge using $\lambda_{\text{GCV}}$.

\section{Extensions}
\subsection{Extension 1: Mixed effects model}\label{sectmixed}
A natural extension of the high-dimensional random effects model (\ref{model}) is the mixed effects model:
\begin{equation}\label{mixed}
\By = \BX_{\text{f}}\boldsymbol{\alpha} + \BX_{\text{r}}\Bbeta + \Beps,
\end{equation}
where we assume that the $n \times m$ design matrix for the fixed effects, $\BX_{\text{f}}$, is of low-rank, so $m \ll n$, as opposed to the random effects design matrix $\BX_{\text{r}}$.
Restricted maximum likelihood (REML) deals with the fixed effects by contrasting them out. For the error contrast vector $\By -  \BX_{\text{f}}\hat{\alpha}^{\text{OLS}}= \BA^T\By,$ with $\BA = \BI_n -  \BX_{\text{f}}(\BX_{\text{f}}^T\BX_{\text{f}})^{-1}\BX_{\text{f}}^T$, the marginal likelihood for the variance components equals (see e.g. \cite{zhang2015tutorial}):

\begin{equation}\label{reml}
P(\BA^T\By) = \mathcal{N}(\By; \boldsymbol{\mu} = \mathbf{0}, \boldsymbol{\Sigma}= \BA^T\boldsymbol{\Sigma}_r\BA)
\end{equation}
with $\boldsymbol{\Sigma}_r = \BX_r \BX_r^{T} \tau^{2} + \sigma^{2} \BI_{n}.$
In addition to maximizing (\ref{reml}) as a function of $(\sigma^2, \tau^2)$, we attempted solving the set of two estimation equations suggested by \cite[]{jiang2007}, but this rendered instable results inferior
to maximizing (\ref{reml}) directly.



\begin{figure}[H]
\includegraphics[width=1\linewidth]{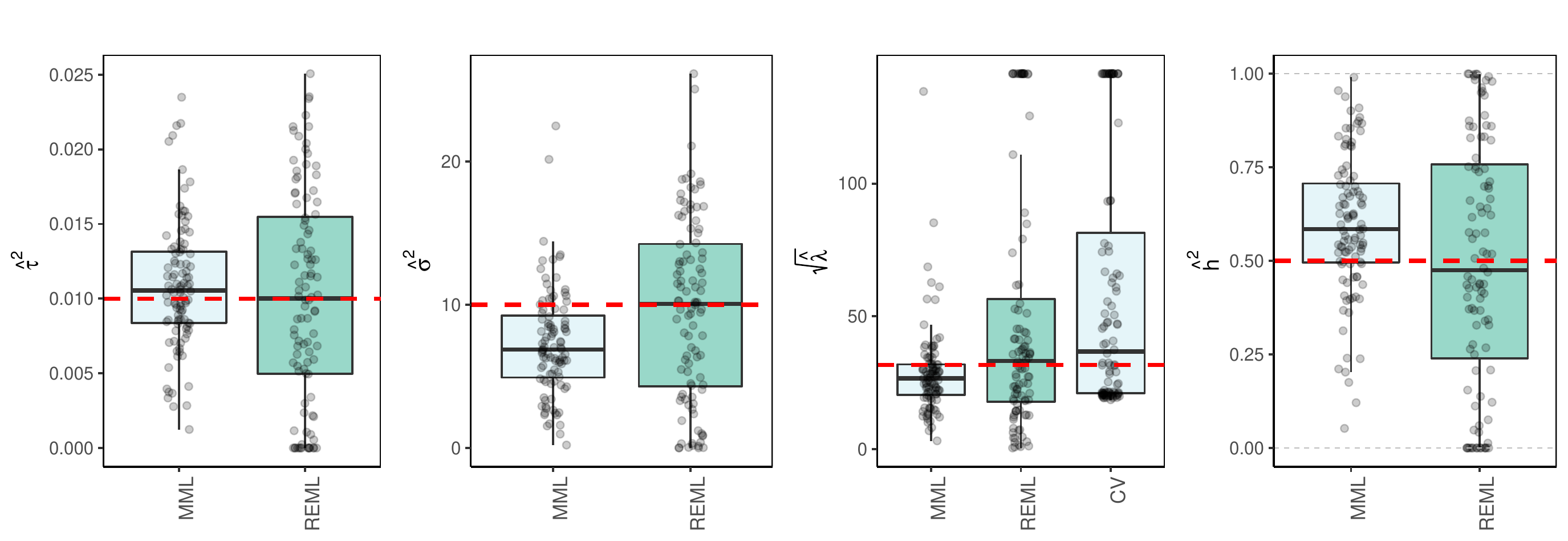}
  \caption{Estimates for mixed effects model, $\tau^{2} = 0.01, \sigma^{2} = 10, n = 100, m = 10, p = 1000$}

 \label{fig:mixedeffects}
\end{figure}

Alternatively, MML may be used, but it has to be adjusted to also estimate the fixed effects in the model. This implies replacing $\boldsymbol{0}$ in Gaussian likelihood \eqref{eq:MML} by
$\BX_{\text{f}}\boldsymbol{\alpha}$, and optimizing \eqref{eq:MML} with respect to $2+m$ parameters, where $m$ is the number of fixed parameters.
The mixed model simulation setting is as follows:
\begin{equation}
\begin{aligned}
&\By_{n \times 1} = \BX_{\text{f},n \times m}\boldsymbol{\alpha}_{m \times 1} + \BX_{\text{r},n \times p}\Bbeta_{p \times 1} + \Beps_{n \times 1} &\qquad &\epsilon_i \iid \mathcal{N}(0, \sigma^{2})\\
&x_{f,ik} \iid \mathcal{N}(0, 1) &\qquad &x_{r,ij} \iid \mathcal{N}(0, 1) \\
&\alpha_{k}  \iid  p_{0,f} \delta_{0} + (1-p_{0,f})\mathcal{N}(0, \tau_{0,f}^2)& \qquad &\beta_j \iid p_{0}\delta_{0} + (1-p_{0})\mathcal{N}(0, \tau_{0}^{2}) ,
\end{aligned}
\end{equation}
where $n = 100, p = 1000, m=10, p_0 = 0.9, \tau_0^2 = 0.1$ (implying variance $\tau^2 = (1-p_0)\tau^2_0=0.01$ for generating random effects) and $p_{0,f}= 0.5, \tau^2_{0,f} = 0.20$ (implying variance $\tau^2_f = 0.1$ for generating fixed effects). Note that we focused on a fairly sparse setting for the random effects and larger prior variance of fixed effects than of random effects, which enables a stronger impact of the small number of fixed effects. Figure \ref{fig:mixedeffects} shows the results of REML,  MML and CV (by \texttt{glmnet}, using penalty factor 0 for the fixed effects) for the estimation of $\tau^2, \sigma^2, \lambda$ and $h^2$.

From Figure \ref{fig:mixedeffects} we observe that REML indeed improves MML in terms of bias, however at the cost of increased variability. For the estimation of $\lambda$, CV is fairly competitive to REML and MML, although it renders markedly more over-penalization.

\subsection{Extension 2: Bayesian linear regression}
\begin{figure}[H]
\includegraphics[width=1\linewidth]{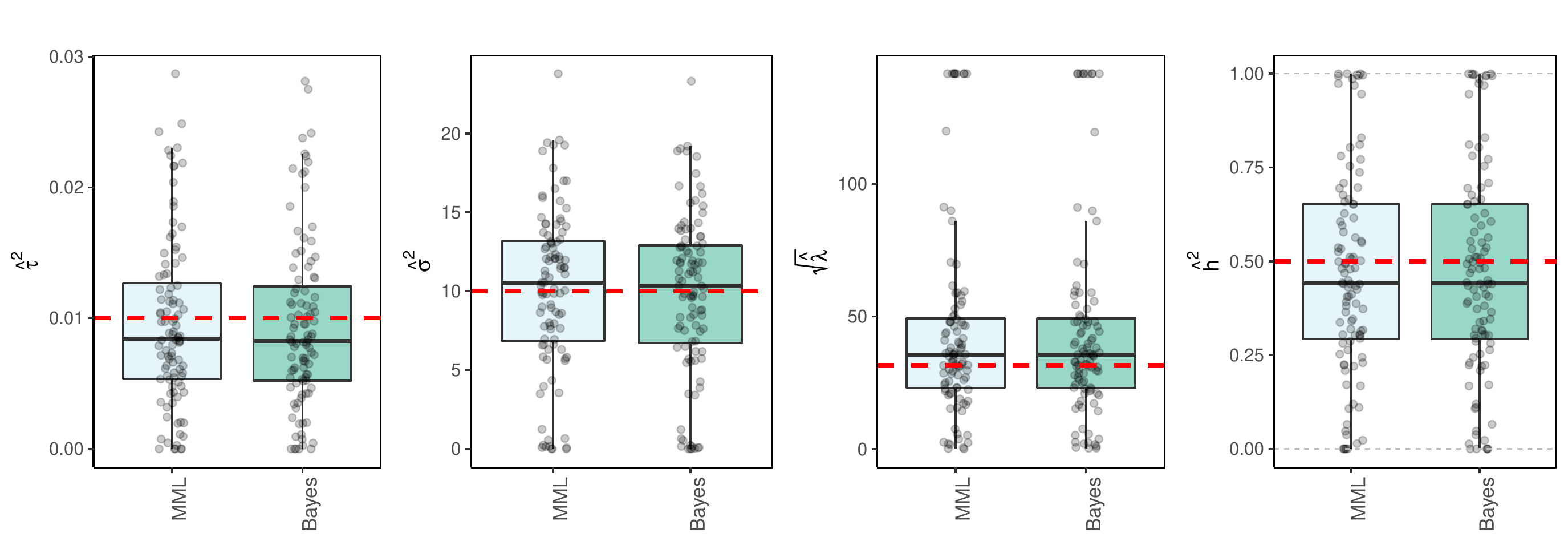}
  \caption{Bayes and MML \eqref{MVN} estimates for multi-collinear $\BX$, with $\tau^{2} = 0.01, \sigma^{2} = 10, n = 100, p = 1000$}

 \label{fig:bayes}
\end{figure}

So far, we focused on classical methods. Bayesian methods may be a good alternative. We applied the standard Bayesian linear regression model, i.e. the conjugate model
with i.i.d. priors $\pi(\beta_j) = N(0,\sigma^2\tau^2)$, with $\tau^2$ fixed and $\sigma^2$ endowed with a vague inverse-gamma prior (see Supplementary Material for details). For this model the maximum marginal likelihood estimator for $\tau^2$ is still analytical \cite{karabatsos2017marginal}, and so is the posterior mode estimate of $\sigma^2$. Figure \ref{fig:bayes} shows the results in comparison to MML, i.e. maximization of \eqref{MVN}, for the random effects case with multi-collinear $\BX$, as in Section \ref{multicol}. Results for other settings were in essence very similar.

From the results we conclude that the conjugate Bayes estimates are very close to those of MML. This is in line with the fact that this conjugate model with prior variance $\tau^2 = \sigma^2/\lambda$ is known to render posterior mean estimates of $\Bbeta$ that equal the $\lambda$-penalized ridge regression estimates.

The conjugate Bayesian model is scale-invariant, because the $\beta$ prior contains the error variance $\sigma^2$. Recently, it was criticized for its non-robustness against misspecification of the fixed $\tau^2$ when estimating $\sigma^2$ \cite{moran2018variance}. However, in practice one needs to estimate $\tau^2$ by either empirical Bayes (e.g. maximum marginal likelihood) or full Bayes. We repeated the simulation by \cite{moran2018variance} (see Supplementary Material). The results show that the estimates of $\sigma^2$ are much better when estimating $\tau^2$ by empirical Bayes instead of fixing it, and in fact very competitive to alternatives proposed by \cite{moran2018variance}.


\subsection{Extension 3: Generalized linear models}
\subsubsection{Setting}
Motivated by the good results for MML in the linear setting, we wish to extend MML estimation to the high-dimensional generalized linear model (GLM) setting, where the likelihood depends on the regression parameter $\bbeta$ only via the linear predictor, $\BX\bbeta$.
Hence, likelihood $\L(\BY; \bbeta,\BX)$ is defined by a density $f_{\mu}(\BY)$ (e.g. Poisson), where $\BX\bbeta$ is mapped to $\mu$ by a link function (e.g. $\log$). As before, we a priori assume i.i.d.
$\beta_j \sim N(0,\tau^2)$, here equivalent to an $L_2$ penalty $\lambda = 1/\tau^2$ when estimating $\bbeta$ by penalized likelihood.
In \cite{Heisterkamp1999} an iterative algorithm to estimate $\lambda$ is derived which alternates estimation of $\bbeta$ by maximization w.r.t. $\lambda$, requiring the computation of the trace of a Hessian of a $p \times p$ matrix. Here, the estimation of $\bbeta$ itself is much slower than in the linear case, because it is not analytic and requires iterative weighted least squares approximation.
Below we show how to substantially alleviate the computational burden in the $p \ggg n$ setting by re-parameterizing the marginal likelihood implying computations in $\mathbb{R}^n$ instead of $\mathbb{R}^p$.

\subsubsection{Method}
We have for the marginal likelihood:
\begin{equation}\label{marglik}
\ML(\lambda) =  \int_{\bbeta \in \mathbb{R}^p} \L(\BY; \bbeta,\BX) \pi_{\lambda}(\bbeta) d\bbeta =\int_{\bbeta \in \mathbb{R}^p} \L(\BY; \bbeta,\BX) \phi(\beta_1;0,1/\lambda) \cdots
\phi(\beta_p;0,1/\lambda)d\bbeta,
\end{equation}
where $\phi(\beta, \mu, \tau^2)$ denotes the normal density with mean $\mu$ and variance $\tau^2$. Now a crucial observation is that for GLM:
\begin{equation}\label{marglik2}
\ML(\lambda) = E_{\pi_{\lambda}(\bbeta)}[\L(\BY; \bbeta,\BX)] = E_{\pi_{\lambda}(\bbeta)}[\L(\BY; \BX\bbeta)] =
E_{\pi'_{\lambda}(\BX\bbeta)}[\L(\BY; \BX\bbeta)],
\end{equation}
because the likelihood depends on $\bbeta$ only via the linear predictor $\BX\bbeta$.
Here, $\pi'_{\lambda}(\BX\bbeta)$ is the implied $n$-dimensional prior distribution of $\BX\bbeta$. This is a multivariate normal: $\boldsymbol{\phi}(\bbeta^{\BX}; \boldsymbol{\mu}=\boldsymbol{0},
\Sigma_\lambda= \BX\BX^T/\lambda)$. Therefore, we have:
\begin{equation}\label{marglik3}
\begin{split}
\ML(\lambda) &= \int_{\bbeta \in \mathbb{R}^p} g_{\BY, \lambda}(\bbeta)d\bbeta = \int_{\bbeta \in \mathbb{R}^p} \L(\BY; \bbeta,\BX) \phi(\beta_1;0,1/\lambda) \cdots  \phi(\beta_p;0,1/\lambda)d\bbeta
\\ &=  \int_{\bbeta^{\BX} \in \mathbb{R}^n} h_{\BY, \lambda}(\bbeta^{\BX})d\bbeta^{\BX}
= \int_{\bbeta^{\BX} \in \mathbb{R}^n} \L(\BY; \bbeta^{\BX},\I_n) \boldsymbol{\phi}(\bbeta^{\BX}; \boldsymbol{0},\Sigma_\lambda) d\bbeta^{\BX}.
\end{split}
\end{equation}
Hence, the $p$-dimensional integral may be replaced by an $n$-dimensional one, with obvious computational advantages when $p \ggg n$. Moreover, the use of (\ref{marglik3}) allows applying implemented Laplace approximations, which tend to be more accurate in lower dimensions.
The Laplace approximation requires $\hat{\bbeta}^{\BX} = \argmax._{\bbeta^{\BX}}\{h_{\BY, \lambda}(\bbeta^{\BX})\}$. We emphasize that this does generally not equal $\BX\hat{\bbeta},$ where $\hat{\bbeta} = \argmax._{\bbeta}\{g_{\BY, \lambda}(\bbeta)\}$:
the maximum of the commonly used $L_2$ penalized (log)-likelihood. However, $\hat{\bbeta}^{\BX}$ can be computed by noting that
\begin{equation}\label{hlaplace}
\log h_{\BY, \lambda}(\bbeta^{\BX}) \propto \ell(\BY; \bbeta^{\BX}, \I_n) -  (\bbeta^{\BX})^T \Sigma_{\lambda}^{-1} \bbeta^{\BX}.
\end{equation} In other words, this is the penalized log-likelihood when regressing $\BY$ on
the identity design matrix $\I_n$ using an $L_2$ smoothing penalty matrix $(\bbeta^{\BX})^T \Sigma_{\lambda}^{-1} \bbeta^{\BX} = \lambda (\bbeta^{\BX})^T (\BX\BX^T)^{-1} \bbeta^{\BX}$. The latter fits conveniently into the set-up
of \cite{wood2011fast}, as implemented in the R-package \texttt{mgcv}. This also facilitates MML estimation of $\lambda$ by maximizing $\ML(\lambda)$, with $h_{\BY, \lambda}(\bbeta^{\BX})$ as in \eqref{hlaplace}. If the columns of $\BX$ are standardized (common in high-dimensional studies), $\BX\BX^T$ has rank $n-1$ instead of $n$, implying that
$(\BX\BX^T)^{-1}$ does not exist and should be replaced by a pseudo-inverse $(\BX\BX^T)^{+}$, such as the Moore-Penrose inverse.

In a full Bayesian linear model setting, dimension reduction is also discussed by \cite{West2003}, where $\BX\bbeta$ is substituted by a $n$-dimensional factor analytic representation, which requires an SVD of $\BX$. In addition, there it is not used for hyper-parameter estimation by marginal likelihood, but instead for specifying (hierarchical) priors for the factors.

\subsubsection{Results}
\begin{figure}[H]
\centering
\begin{subfigure}{.5\textwidth}
  \centering
 \caption{Poisson for independent $\BX$}
  \includegraphics[scale= 0.5]{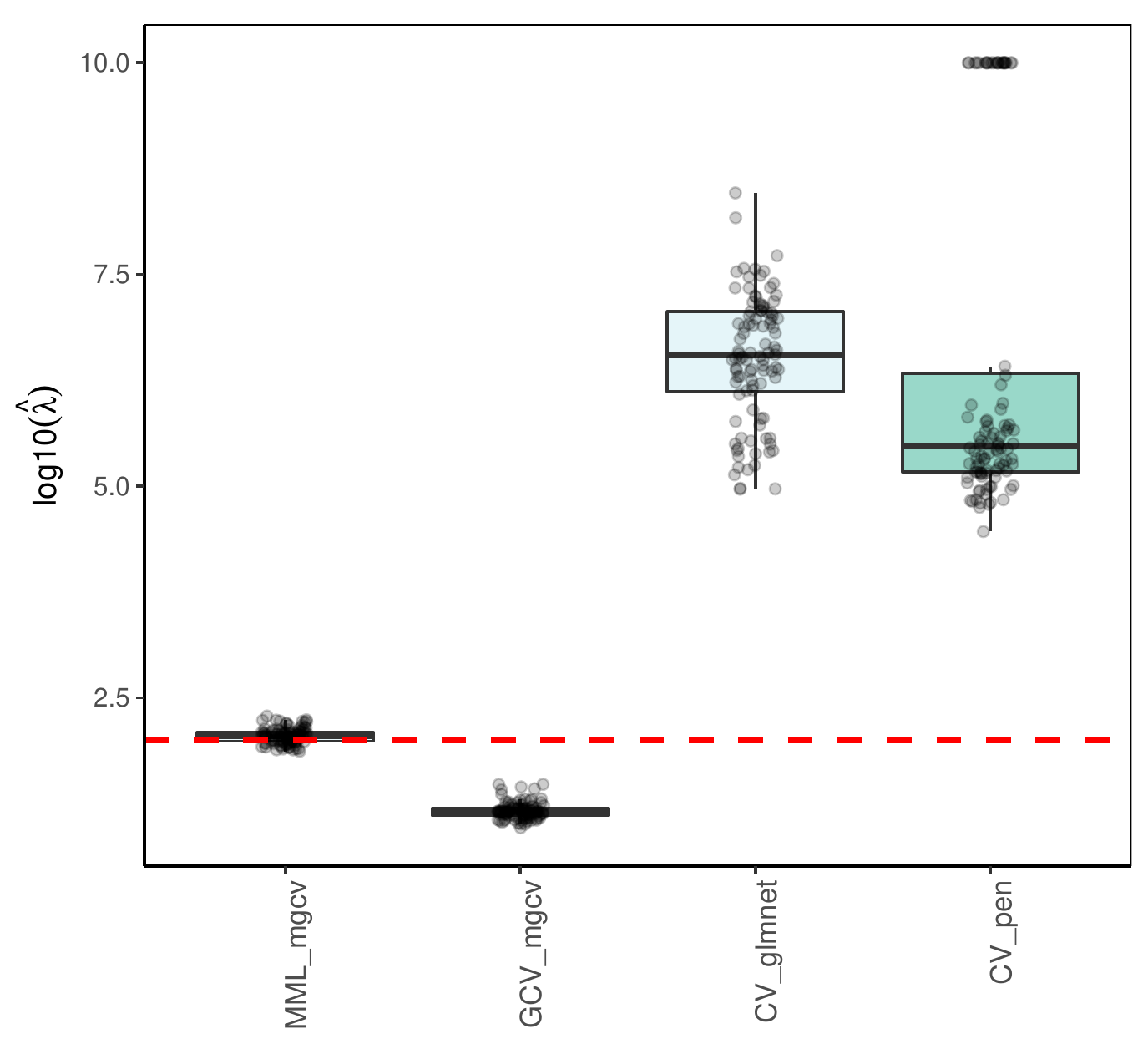}

\end{subfigure}%
\begin{subfigure}{.5\textwidth}
  \centering
   \caption{Poisson for multi-collinear $\BX$}
  \includegraphics[scale= 0.5]{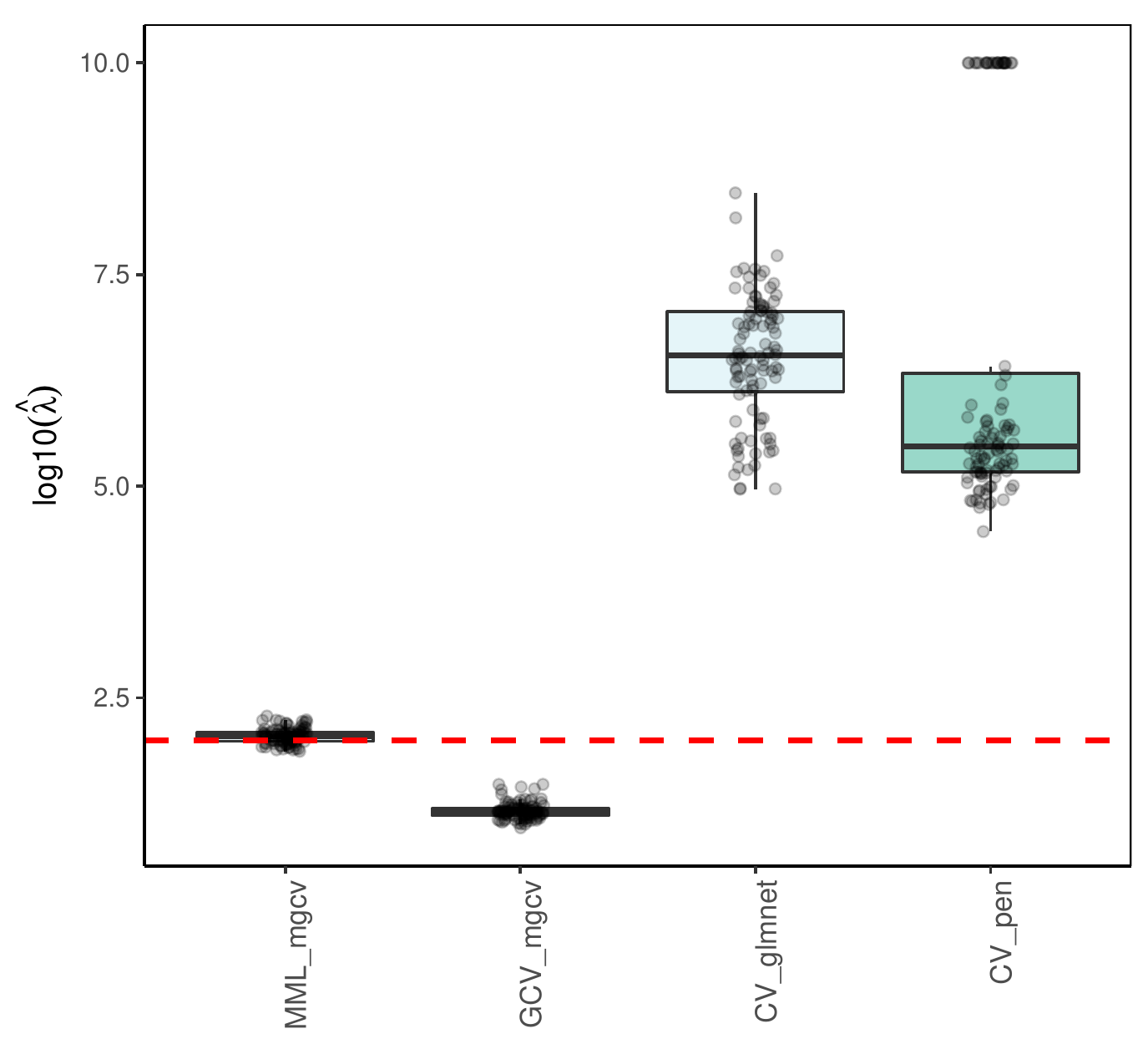}
  \label{GCV_Herit_ServerHighDim}
\end{subfigure}
\caption{$\lambda$ estimates for Poisson ridge regression, $\lambda = 1/\tau^{2} = 100, n = 100, p = 1000$.}
\label{fig:GLM}
\end{figure}

R packages like \texttt{glmnet} \cite{Friedman2010} and \texttt{penalized} \cite{Goeman2010b} estimate $\lambda$ by cross-validation, and also \texttt{mgcv} allows, next to the MML estimation, (generalized) CV estimation \cite[]{wood2011fast}.
Figures \ref{fig:GLM}(a)and \ref{fig:GLM})(b) show the results for Poisson ridge regression, with $\BY_i \sim \text{Pois}(\lambda_i), \lambda_i = \exp(\BX_i \bbeta)$,
$\bbeta$ generated as in \eqref{basicmodel}, and $\BX$  generated as in \eqref{basicmodel} and \eqref{multicoll}, which denote the independent $\BX$ and multi-collinear $\BX$ setting, respectively.

Figure \ref{fig:GLM} clearly shows the superior performance of MML based on \eqref{marglik3} over CV. In particular, \texttt{glmnet} and \texttt{penalized} render strongly upward biased values.
The \texttt{mgcv} GCV values are still inferior to MML based ones, but much better than the latter two, which may be due to the different regression estimators used (Laplace approximation versus iterative weighted least squares).
 We should stress that CV does not target for the estimation of $\lambda$ as such, but merely for minimizing prediction error. Nevertheless, the difference is remarkably larger than in the corresponding linear case (see Figures \ref{fig:indX} and \ref{fig:Cor}).

The Supplementary Material shows the results for Binomial ridge regression. While the differences in performance are less dramatic than for the Poisson setting,
MML still renders much better estimates of $\lambda$ than CV-based approaches.

\section{Computational aspects and software}
All methods and simulations presented here are implemented in a few wrapper \texttt{R} scripts: one for the linear random effects model (which includes the conjugate Bayes estimator), one for the linear mixed effects model, and
one for Poisson and Binomial ridge regression. Parallel computations are supported. The scripts allow exact reproduction of the results in this manuscript as well as comparisons for other simulation or user-specific real data $\BX$ cases. In addition, a script is supplied to produce the box-plots as in this manuscript.

HiLMM, PCR and CV implementations are provided by the R-packages \texttt{HiLMM, v1.1} \cite{BonnetHiLMM}, \texttt{ridge, v1.8-16} \cite{Cule2012} (code slightly adapted for computational efficiency) and \texttt{glmnet, v2.0-16} \cite{Friedman2010}.
The methods MML, REML, Bayes, MoM, Basic and GCV were implemented by us for the linear random and mixed effects models. For Poisson and Binomial ridge regression we applied  \texttt{mgcv, v1.8-16} \cite[]{wood2011fast} after our re-parametrization \eqref{marglik3} to obtain MML and GCV results, while for CV  \texttt{glmnet} and \texttt{penalized, v0.9-50} \cite{Goeman2010b} were applied. For all methods that required optimization the \texttt{R} routine \texttt{optim} was used, with default settings. CV was based on 10 folds.

Computing times of the various methods largely depend on $n$ and $p$, much less so on the exact simulation setting. These are displayed for $n=100, 500$ and $p= 10^3, 10^4, 10^5$ in Table \ref{tab:ct}, based on computations with
one CPU of an \texttt{Intel\textsuperscript{\textregistered} Xeon\textsuperscript{\textregistered} CPU E5-2660 v3 @ 2.60GHz} server.
For Poisson ridge regression, we only report the computing times of MML and GCV, because, as reported in Figure \ref{fig:GLM}, the performance of CV-based methods was very inferior.
\begin{table}[ht]
\centering
{\renewcommand{\arraystretch}{0.7}
\begin{tabular}{|r|rrr||rrr|}
  \hline
 \textbf{Linear}  & \multicolumn{3}{|c||}{$n=100$} & \multicolumn{3}{c|}{$n=500$}\\
 & $p= 10^3$ & $p= 10^4$ & $p= 10^5$ & $p= 10^3$ & $p= 10^4$ & $p= 10^5$ \\
  \hline
MML & 0.06 & 0.15 & 1.12 & 2.18 & 6.07 & 26.64 \\
  Bayes & 0.04 & 0.31 & 4.38 & 1.10 & 7.78 & 93.25 \\
  MoM & 0.01 & 0.08 & 1.03 & 0.17 & 2.32 & 23.70 \\
  PCR & 0.05 & 0.39 & 5.36 & 1.39 & 10.31 & 116.80 \\
  Basic & 0.05 & 0.46 & 6.56 & 1.44 & 12.40 & 145.18 \\
  GCV & 0.20 & 0.46 & 4.56 & 12.26 & 26.41 & 111.38 \\
  CV & 0.81 & 6.57 & 39.95 & 2.62 & 21.69 & 183.50 \\
  HiLMM & 0.03 & 0.17 & 2.01 & 0.66 & 3.14 & 27.99 \\
   \hline
  \hline
\textbf{ Poisson}  & \multicolumn{3}{|c||}{$n=100$} & \multicolumn{3}{c|}{$n=500$}\\
 & $p= 10^3$ & $p= 10^4$ & $p= 10^5$ & $p= 10^3$ & $p= 10^4$ & $p= 10^5$ \\
 \hline
MML\_mgcv & 0.32 & 0.33 & 0.31 & 26.21 & 40.19 & 48.17 \\
GCV\_mgcv & 0.39 & 0.33 & 0.62 & 33.48 & 41.44 & 54.01 \\
   \hline
\end{tabular}\caption{Computing times for hyper-parameter estimation for linear and Poisson ridge regression}
}
\end{table}\label{tab:ct}
From Table \ref{tab:ct} we conclude that MML is also computationally very attractive. Its efficiency is explained by the fact that, unlike many of other methods, it does not require an SVD or other matrix decomposition of $\BX$. Moreover, the only computation that involves dimension $p$ is the product $\BX \BX^T$.




\section{Discussion}
We compared several estimators in a large variety of high-dimensional settings. The results showed that plain maximum marginal likelihood works well in many settings.
MML is generally superior to methods that aim to separately estimate $\sigma^2$ (\ref{MSE}, \ref{pcr}). Apparently, the estimates of $\sigma^2$ and $\tau^2$ are so intrinsically linked in the high-dimensional setting that separate estimation is sub-optimal. The moment estimator (MoM) is generally not competitive to MML. It may, however, be useful in large systems with multiple hyper-parameters to estimate \emph{relative} penalties, which
are less sensitive to scaling issues than the global penalty parameter \cite[]{wielGRridge}. MoM may also be a useful initial estimator for more complex estimators that are based on optimization, such as MML.

Possibly somewhat surprising is the good performance of MML for estimating $\lambda$ and $h^2$, as these are functions of $\sigma^2$ and $\tau^2$. For the estimation of $\lambda$ it is generally better than or competitive to (generalized) CV, an observation also made for the low-dimensional setting \cite{wood2011fast}. The inferior performance of the basic estimator of $\sigma^2$ (\ref{MSE}) implies that alternative estimators of $\lambda$ that use $\hat{\sigma}^2$ as a plug-in are unlikely to perform well in high-dimensional settings. Such estimators, including the original one by Hoerl and Kennard \cite{Hoerl1970}, are compared by \cite{Muniz2009, kibria2016some}, who show that some do perform well in the \emph{low-dimensional} setting. For Poisson ridge regression, similar estimators of $\lambda$ are available \cite{maansson2011poisson}, but these rely on an initial maximum likelihood estimator of $\bbeta$, and hence do not apply to the high-dimensional setting. For estimating $h^2$ it should be noticed that \texttt{HiLMM} \cite{BonnetHiLMM} aims to compute a confidence interval for $h^2$ as well. For that purpose their direct estimator (\ref{h2}) is likely more useful than MML on the pair $(\tau^2, \sigma^2)$. We also used \texttt{Esther} \cite{BonnetEsther}, which precedes \texttt{HiLMM} by sure independence screening. It did not improve \texttt{HiLMM} in our (semi-)sparse settings, and requires manual steps. However, it likely improves \texttt{HiLMM} results in very sparse settings \cite{BonnetEsther}.

For mixed effect models with a small number of fixed effects, MML compares fairly well to REML, with a larger bias, but smaller variance. Probably the potential advantage of contrasting out the fixed effects is small when the number of random effects is large. REML may have a larger advantage in very sparse settings \cite{Jiang2016} or when the number of fixed effects is large with respect to $n$.
Estimates from the conjugate Bayes model are very similar to those by MML. We show that estimating $\tau^2$ along with $\sigma^2$ highly improves the $\sigma^2$ estimates presented by \cite{moran2018variance},
where a fixed value of $\tau^2$ is used. In the case of many variance components or multiple similar regression equations, Bayesian extensions that shrink the estimates by a common prior are appealing, in particular in combination with efficient posterior approximations such as variational Bayes \cite{Leday2017}.

Our model (\ref{model}) implies a dense setting, but we have demonstrated that the MML and REML estimators of $\tau^2$ and $\sigma^2$ are fairly robust against moderate sparsity, which corroborates the results by \cite{Jiang2016}. Nevertheless, true sparse models may be preferable when variable selection is desired, which depends on accurate estimation of $\bbeta$. On the other hand, post-hoc selection procedures can be rather competitive \cite{Bondell2012}. Moreover, the sparsity assumption is questionable for several applications. E.g. in genetics, it was suggested that many complex traits (such as height or cholesterol levels) are not even polygenic, but instead ``omnigenic'' \cite{boyle2017expanded}.

The extension of MML to high-dimensional GLM settings \eqref{marglik3} is promising given its computational efficiency and performance for Poisson and Binomial regression. A special case of the latter, logistic regression,
requires further research, because the Laplace approximations of the marginal likelihood are less accurate here \cite{wood2011fast}. Extension to survival is a promising avenue, because
Cox regression is directly linked to Poisson regression \cite{cai2003hazard}. Alternatively, parametric survival models may be pursued. To what extent the estimates of hyper-parameters impact predictions depends on the sensitivity of the likelihood to these parameters. For the linear setting, a re-analysis of the weight-gain data showed that predictions based on $\hat{\lambda}_{\text{MML}}$ improved those based on $\hat{\lambda}_{\text{CV}}$.


The MML estimator can be extended to estimation of multiple variance components or penalty parameters, which was addressed by iterative likelihood minorization \cite{zhou2015mm} and by
parameter-based moment estimation \cite[]{wielGRridge}. The latter extends to non-Gaussian response such as survival or binary.
Further comparison of these methods with multi-parameter MML, both in terms of performance and computational efficiency, is left for future research. Finally, in particular in genetics applications, extensions of estimation of variance components by MML to non-independent individuals can be implemented by use of a well-structured between-individual covariance matrix $\boldsymbol{\Sigma}$ \cite{kang2008efficient}.

Although our simulations cover a fairly broad spectrum of settings, many other variations could be of interest. We therefore supply fully annotated \texttt{R} scripts \url{https://github.com/markvdwiel/Hyperpar} that allow i) comparison of all algorithms discussed here,  also for one's `own' real covariate set $\BX$; and ii) reproduction of all results presented here.

\section*{Acknowledgement} Gwena\"el Leday was supported by the Medical Research Council, grant number MR/M004421. We thank Jiming Jiang and Can Yang for their correspondence, input and software for the MM algorithm. In addition, we thank Kristoffer Hellton for providing the \texttt{fridge} software and data. Finally, Iuliana Cioc\v anea-Teodorescu is acknowledged for preparing the TCGA KIRC data.


\section*{Supplementary Material}
\subsection{Contents}
This Supplementary Material contains:
\begin{itemize}
  \item Details on the estimation of $\sigma^2$ and $\tau^2$ with the conjugate Bayesian model
  \item Estimation results from this Bayesian model for a simulation by \cite{moran2018variance}
  \item Details on the TCGA KIRC and TCPA OV gene and protein expression data
  \item An alternative analysis of the Weight gain data
  \item Supplementary Figures:
  \begin{itemize}
  \item Results for $\lambda$ estimation for Binomial ridge regression
  \item Robustness of estimates for non-Gaussian $\beta$'s and errors
  \end{itemize}
\end{itemize}

\subsection{Bayesian linear regression}
\subsubsection{Method}
The conjugate Bayesian linear regression model is:
\begin{equation}
	\label{bmodel}
	\begin{split}
	\By_{n\times 1} &= \BX_{n\times p}\Bbeta_{p\times 1} + \Beps_{n\times 1},  \\
	\Bbeta &\sim \mathcal{N}(0, \sigma^2 \tau^2 I_p), \\
	\Beps &\sim \mathcal{N}(0,\sigma^2 I_n),\\
	\sigma^{-2} &\sim \mathcal{G}(a,b)
	\end{split}
\end{equation}
\null
where $\BX$ is the design matrix, $\Bbeta$ is the vector of unknown regression parameters, $\epsilon$ is the vector of random errors and the prior shape and rate hyper-parameters $a$ and $b$ are fixed (say $a=1$ and $b=0.001$) so as to induce a non-informative prior. In model (\ref{bmodel}) the ridge regularization parameter corresponds to $\nu=\tau^{-2}$. In the following we provide the marginal posterior distribution of $\Bbeta$ and $\sigma^{-2}$, as well as a reformulation of the marginal likelihood \cite{karabatsos2017marginal} that allows important computational savings.

The likelihood function and prior densities are
\begin{equation*}
	\begin{split}
		\pi(\By|\Bbeta,\sigma^{-2}) &= \displaystyle{ (2\pi)^{-n/2}(\sigma^{-2})^{n/2} \text{exp}\left\lbrace -\frac{1}{2}\sigma^{-2}(\By-\BX\Bbeta)^T (\By-\BX\Bbeta) \right\rbrace}\\
		\pi(\Bbeta|\sigma^{-2}, \tau^{-2} )	&= \displaystyle{ (2\pi)^{-p/2}(\sigma^{-2}\tau^{-2})^{p/2} \text{exp}\left\lbrace -\frac{1}{2} \sigma^{-2}\tau^{-2} \Bbeta^T \Bbeta \right\rbrace}\\
		\pi(\sigma^{-2}) 	&= \displaystyle{ \frac{b^a}{\Gamma\left(a\right)} (\sigma^{-2})^{a-1} \text{exp}\left\lbrace-b \sigma^{-2} \right\rbrace}.
	\end{split}
\end{equation*}
\null

\noindent Therefore, the joint posterior is given by\\[10pt]
$\pi(\By|\Bbeta,\sigma^{-2})\pi(\Bbeta|\sigma^{-2}, \tau^{-2} )\pi(\sigma^{-2}) = $\\[5pt]
\null\hfill$\displaystyle{(2\pi)^{-\frac{n+p}{2}} \frac{b^a}{\Gamma\left(a\right)} (\sigma^{-2})^{\frac{n+p+2a+2}{2}} (\tau^{-2})^{\frac{p}{2}} \text{exp}\left\lbrace -\frac{1}{2}\sigma^{-2}\left[(\mathbf{y}-\BX\Bbeta)^T (\By-\BX\Bbeta) + \tau^{-2}\Bbeta^T \Bbeta + 2b \right]\right\rbrace }$\\[10pt]

\noindent The marginal posterior distribution of $\Bbeta$ is recognized to be a Student distribution with $2a+n$ degrees of freedom:
\begin{equation*}
		\pi(\Bbeta|\By) = \int \pi(\By|\Bbeta,\sigma^{-2})\pi(\Bbeta|\sigma^{-2}, \tau^{-2} )\pi(\sigma^{-2})\ d\sigma^{-2} =^d \mathcal{T}_{2a+n}(\Bbeta^\ast_\nu, \Sigma^\ast_\nu).\\
\end{equation*}
\null
Here $\Bbeta^\ast_\nu=V^\ast_\nu \BX^Ty$, $\displaystyle{\Sigma^\ast_\nu= \left(\frac{b^\ast_\nu}{a^\ast}\right)V^\ast_\nu}$ and $V^\ast_\nu=\left(\BX^T\BX+\nu\boldsymbol{\mathbb{I}}_p \right)^{-1}$.
The marginal posterior distribution of $\sigma^{-2}$ is a Gamma distribution:
\begin{equation}
		\pi(\sigma^{-2}|\By) = \int \pi(\By|\Bbeta,\sigma^{-2})\pi(\Bbeta|\sigma^{-2}, \tau^{-2} )\pi(\sigma^{-2})\ d\Bbeta
			=^d \mathcal{G}(a^\ast, b^\ast_{\nu})
\end{equation}\label{pmeanprec}
\null
where $a^\ast=a+0.5 n$ and $b^\ast_{\nu}=\displaystyle{b + 0.5 \left(\By^T\By-{\Bbeta^\ast_\nu}^T {\Sigma^\ast_\nu}^{-1}\Bbeta^\ast_\nu\right)}$
\noindent The marginal likelihood of the model is:
\begin{equation}
	\label{ML}
		\pi(\By) = \iint \pi(\By|\Bbeta,\sigma^{-2})\pi(\Bbeta|\sigma^{-2}, \tau^{-2} )\pi(\sigma^{-2})\ d\Bbeta d\sigma^{-2}
		 	= \frac{\vert\Sigma^\ast_\nu\vert^{1/2} b^a \Gamma(a^\ast)}{\vert\nu^{-1}\mathbb{I}_p\vert^{1/2} {b^\ast_{\nu}}^{a^\ast}\Gamma(a)\pi^{n/2}}
\end{equation}
\null

\indent The marginal likelihood in \eqref{ML} involves the determination of $\vert\Sigma^\ast_\nu\vert$, which can be computationally demanding when the number of variables $p$ is large and when we wish to evaluate the marginal likelihood for various values of $\nu$. To tackle this problem it is helpful to consider the singular value decomposition $UDV^T$ of $\BX$, (where $U$ and $V$ are respectively $n\times q$ and $p\times q$ orthogonal matrices and $D=\text{diag}(d_1, \ldots, d_q)$ is the diagonal matrix of singular values $d_1 > \ldots > d_q$ with $q=\text{min}(n,p)$) and focus on the linear model $\By =^d \mathcal{N}(F\Btheta, \sigma^2 \mathbb{I}_n)$, where $F=UD$ and $\Btheta=V^T\Bbeta$, instead of $y =^d \mathcal{N}(\BX\Bbeta, \sigma^2 \mathbb{I}_n)$. Using simple algebra it can be shown that $\mathbb{E}\left[ \Btheta \vert \By \right]=\Btheta^\ast_\nu=V^T\Bbeta^\ast_\nu$
and $\mathbb{V}\left[ \Btheta \vert \By \right]=V^T\Sigma^\ast_\nu V=\left( D^2 + \nu\mathbb{I}_q \right)^{-1}$, which suggest that the marginal likelihood (that is invariant to linear transformations) is more easily determined in the orthogonalized space. Indeed, we now have that
\begin{equation}
	\label{ML2}
	\begin{split}
		\pi(\By) &= \displaystyle{\frac{\vert\left( D^2 + \nu\mathbb{I}_q \right)^{-1}\vert^{1/2} b^a \Gamma(a^\ast)}{\vert\nu^{-1}\mathbb{I}_q\vert^{1/2} {b^\ast_{\nu}}^{a^\ast}\Gamma(a)\pi^{n/2}}} = \displaystyle{\frac{\nu^{q/2} b^a \Gamma(a^\ast)}{\left[\prod_{k=1}^{q}{\left(d_k^2+\nu\right)}\right]^{1/2} {b^\ast_{\nu}}^{a^\ast}\Gamma(a)\pi^{n/2}}}.
	\end{split}
\end{equation}
Additionally, using the fact that
\begin{equation}
	\Btheta^\ast_\nu = \left( \mathbb{I}_q + \nu D^{-2} \right)^{-1} \hat{\Btheta}
\end{equation}
where $\hat{\Btheta} = \left( F^TF \right)^{-1} F^T y = D^{-1} U^Ty$, we deduce that
\begin{equation}
	{\Bbeta^\ast_\nu}^T {\Sigma^\ast_\nu}^{-1}\Bbeta^\ast_\nu = {\Btheta^\ast_\nu}^T\left( D^2 +\nu\boldsymbol{\mathbb{I}}_p  \right) \Btheta^\ast_\nu
	= \hat{\Btheta}^T \left( \mathbb{I}_q + \nu D^{-2} \right)^{-1} D^2 \hat{\Btheta}
	= \displaystyle{\sum_{k=1}^{q}{\frac{\hat{\Btheta}_k^2 d_k^4}{d_k^2+\nu}}}
\end{equation}
and
\begin{equation}
	b^\ast_{\nu} = \displaystyle{b + 0.5 \left(\By^T\By-\sum_{k=1}^{q}{\frac{\hat{\Btheta}_k^2 d_k^4}{d_k^2+\nu}}\right)}.
\end{equation}

\noindent Finally, the log-marginal likelihood reduces to
\begin{equation}
	\label{logML}
		\log \pi(\By) =
		\displaystyle{\frac{q}{2}\log\nu - \frac{1}{2}\sum_{k=1}^{q}{\log\left( d_k^2 + \nu\right)}} - \displaystyle{a^\ast \log\left[ b + 0.5 \left(\By^T\By-\sum_{k=1}^{q}{\frac{\hat{\Btheta}_k^2 d_k^4}{d_k^2+\nu}}\right) \right] + C}
\end{equation}
with constant $\displaystyle{C=a\log b+\log\Gamma(a^\ast) -\log\Gamma(a)-\frac{n}{2}\log\pi}$. Expression \eqref{logML} makes the evaluation of $\log \pi(\By)$ very efficient for different values of $\nu$. Furthermore, the function $\log \pi(\By)$ is $\log$-concave in $\nu$, which facilitates the use of numerical methods over grid-search approaches. In practice, it might be good to employ a numerical algorithm on a sub-domain of $\mathbb{R}^+$ obtained from a rough grid-search.

\subsubsection{Results for simulation by \cite{moran2018variance}}
The $\Bbeta$ prior in the conjugate Bayesian model \eqref{bmodel} is scale-invariant, because it contains the error variance $\sigma^2$. Recently, it was criticized in \cite{moran2018variance} for its failure to estimate $\sigma^2$ well when$\tau^2$ is fixed and misspecified. In reality, $\tau^2$ is often estimated as well, which may improve results. To study this, we repeated the linear regression simulation by \cite{moran2018variance}, which has the following specifications:
$n=100, p = 90,$ with $(\beta_1, \ldots, \beta_6) = (-2.5,-2,-1.5,1.5,2,2.5)$, and $\beta_7 = \beta_8 = \ldots \beta_{90} = 0.$ Moreover $\BX$ was generated from the independent standard Normal, and error variance $\sigma^2 = 3.$

\begin{figure}
\includegraphics{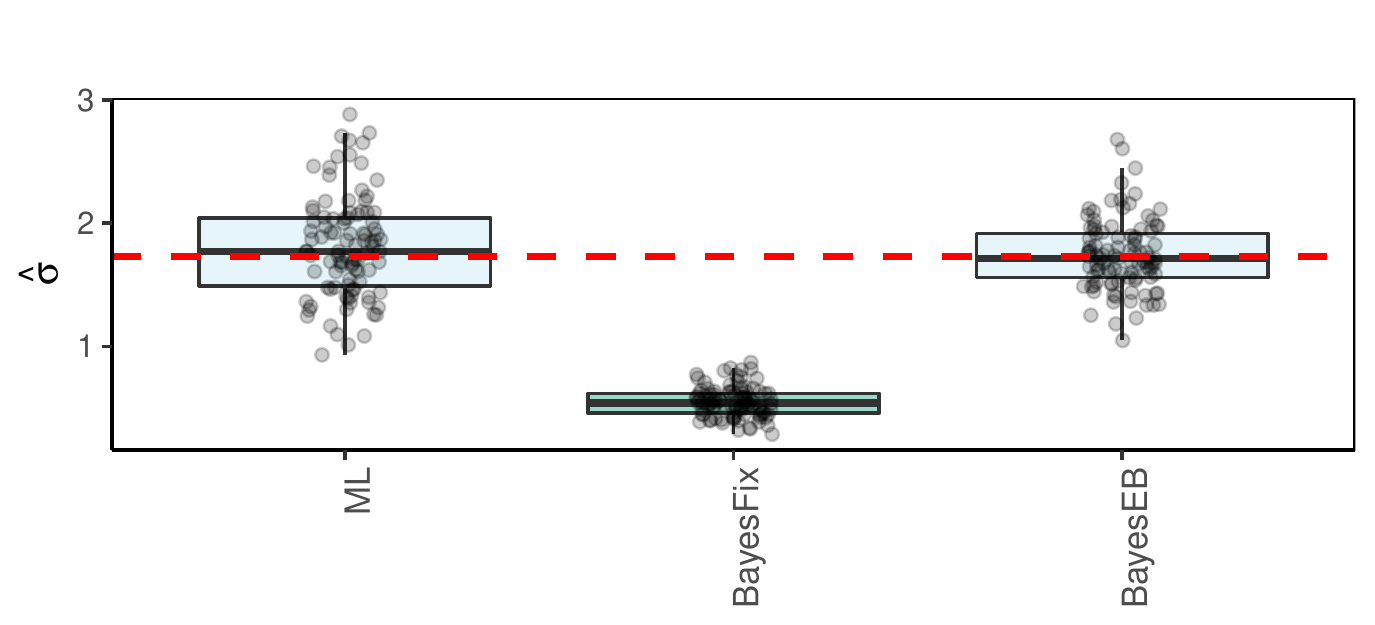}
  \label{simMoran}
\caption{Estimates of $\sigma=\sqrt{3}$ for the simulation by \cite{moran2018variance}}
\label{fig:simMoran}
\end{figure}

Figure \ref{fig:simMoran} shows the results for three methods. Here, \texttt{ML} serves as a benchmark and refers the ordinary maximum likelihood estimator of $\sigma^2$, referred to by \cite{moran2018variance} as `Least Squares'. In addition,
\texttt{BayesEB} and \texttt{BayesFix} estimate $\sigma$ by $(b^\ast_{\nu}/(a^\ast-1))^{1/2},$ i.e. the square-root of the posterior mean of $\sigma^2 = 1/\sigma^{-2}$ \eqref{pmeanprec}, with $\nu = 1/\tau^2$ estimated by empirical Bayes (EB; maximizing \eqref{logML}) and with $\nu$ fixed to $1/100$ (as in \cite{moran2018variance}), respectively.

As in \cite{moran2018variance}, the results show that indeed the conjugate Bayes model is not robust against wrongly fixing $\tau^2$. However, it also shows that EB estimation of $\tau^2$ strongly improves the $\sigma$ estimate. In fact, these estimates are very competitive to the ones for the scale-independent prior, $\beta \sim \mathcal{N}(0, \tau^2 I_p)$, which was advocated by \cite{moran2018variance}.

\subsection{Details on real data used for simulations}
\subsubsection{Kidney tumor gene expression}
The TCGA KIRC data is downloaded from the harmonised GDC database \url{http://cancergenome.nih.gov/} using the R-package TCGAbiolinks \cite{colaprico2015tcgabiolinks}. It contains RNAseq profiles of $n=71$ kidney renal clear cell carcinomas. Only those genes were retained that had more than two counts per million in at least three samples, rendering $p=18,391$ genes.
The data were normalised using the trimmed mean of M-values method of the R-package edgeR \cite{Robinson2010}. Following common practice, all gene expression values were standardized.

\subsubsection{Ovarian tumor protein expression}
We use protein expression data from the cancer proteome atlas (TCPA; \cite{li2013tcpa}), which holds 408 ovarian serous cystadenocarcinoma profiles measuring 224 proteins by reverse-phase protein arrays. These are available from the TCPA portal: \url{https://www.tcpaportal.org/tcpa/}. Data were normalized by median centering per sample. All protein expression values were standardized.


\subsection{Weight gain data example: alternative analysis}
In the main document we re-analysed the weight gain data recently discussed in \cite{hellton2018fridge}. The authors kindly provided the \texttt{R}-code corresponding to the results in \cite{hellton2018fridge},
 which includes their focused ridge (\texttt{fridge}) methodology. Below we report a small discrepancy between our analysis and theirs, and also present results of an alternative analysis.

\begin{figure}
\centering
\includegraphics[scale=0.7]{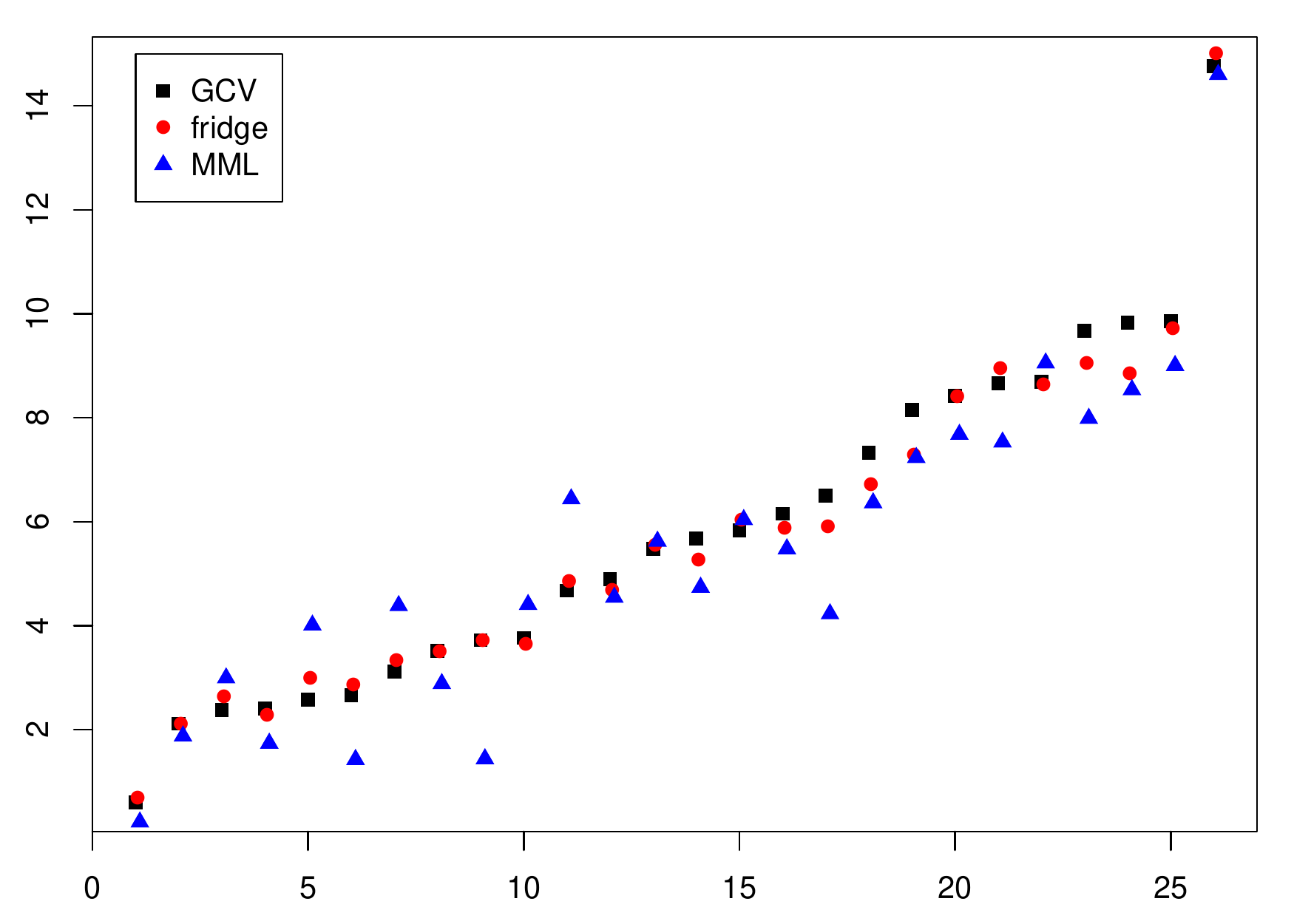}
\caption{Absolute prediction errors (obtained by loocv; y-axis) for ridge using $\lambda_{\text{GCV}}$, for \texttt{fridge} and for ridge using $\lambda_{\text{MML}}$. Sample indices (x-axis) are sorted by GCV results.}
\label{weightgain_cv}
\end{figure}

The authors of \cite{hellton2018fridge} opted to present results on $n=25$ (rather than $n=26$) samples, for technical reasons (personal communication). Results differed very little, so we chose to use all $n=26$ samples of the original study \cite{cashion2013expression}. In addition, the \texttt{R}-code by \cite{hellton2018fridge} includes a prior selection of those 1,000 genes with the largest absolute marginal correlation correlation to the response, i.e. weight gain. Such a prior selection can indeed enhance performance of ridge-type predictors. The analysis by \cite{hellton2018fridge}, however, did not include the gene selection as part of the outer leave-one-out cross-validation loop for assessing predictive performance. This may lead to over-optimism of the prediction errors. We therefore repeated the analysis and comparison as presented in the main document, but with the gene selection
as part of the outer CV-loop. Indeed the error estimates increased substantially: $\text{MSPE}_\text{MML} = 38.65$ (was 14.40), $\text{MSPE}_\text{GCV} = 44.20\ (16.38)$ and $\text{MSPE}_\text{fridge} = 41.65\ (15.80).$
Nevertheless, the conclusion remains that the MML and \texttt{fridge} MSPEs are lower that those of GCV, with relative decreases of 12.5\% and 5.8\%, respectively. Figure \ref{weightgain_cv} displays absolute prediction errors per sample and illustrates the improved prediction by ridge using $\lambda_{\text{MML}}$ with respect to ridge using $\lambda_{\text{GCV}}$.

\subsection{Supplementary Figures}

\subsubsection{Robustness: non-Gaussian $\beta$'s and errors }

Here, we show additional results for the simulation settings with either non-Gaussian $\beta$'s (spike-and-slab or Uniform) or errors ($t_4$), as presented in the main document.
In all cases, parameters of the $\beta$ prior or the error ($\epsilon_i$) distribution were set such that $\tau^2 = \mathbb{V}(\beta_j) = 0.01$ and $\sigma^2 = \mathbb{V}(\epsilon_i) = 10$.

\begin{figure}[H]
\caption{Robustness against non-Gaussian $\beta$'s or errors: hyper-parameter estimates for $n=100, p=1000, \tau^2 = 0.01, \sigma^{2} = 10$}
\centering
\begin{subfigure}{\textwidth}
  \centering
  \includegraphics[width=1\linewidth]{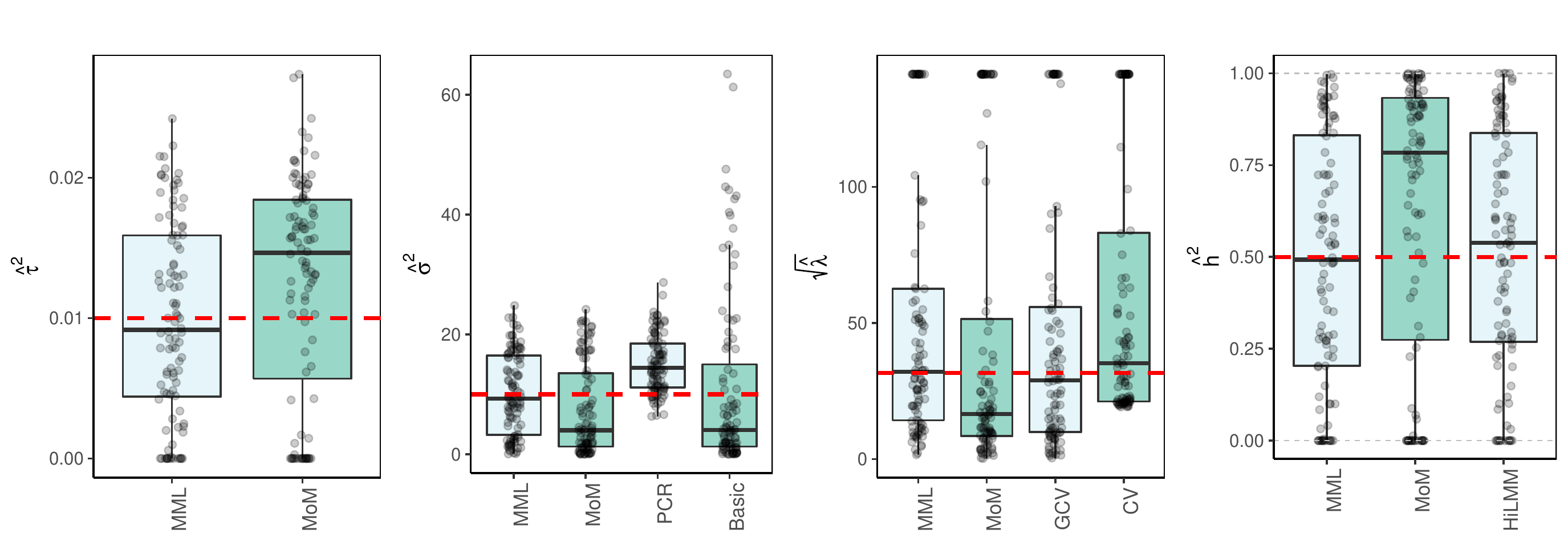}
  \caption{Spike-and-slab $\beta$'s}
\end{subfigure}%

\begin{subfigure}{\textwidth}
  \centering
  \includegraphics[width=1\linewidth]{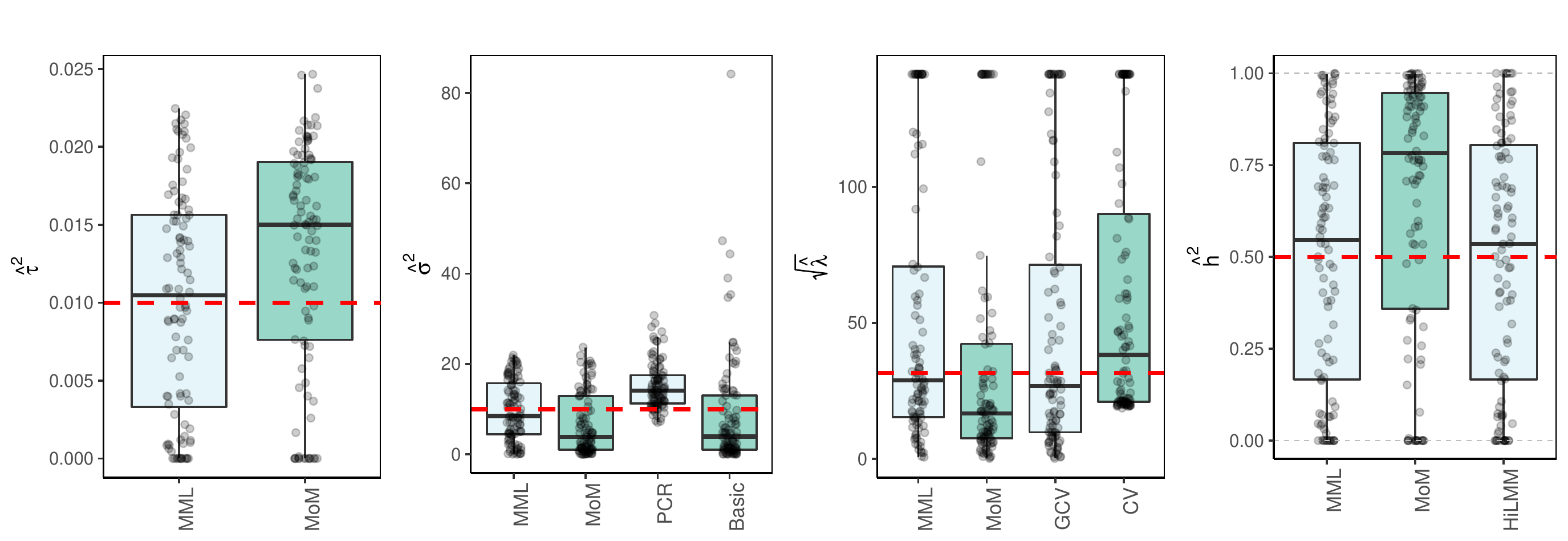}
  \caption{Uniform $\beta$'s}
\end{subfigure}

\begin{subfigure}{\textwidth}
  \centering
  \includegraphics[width=1\linewidth]{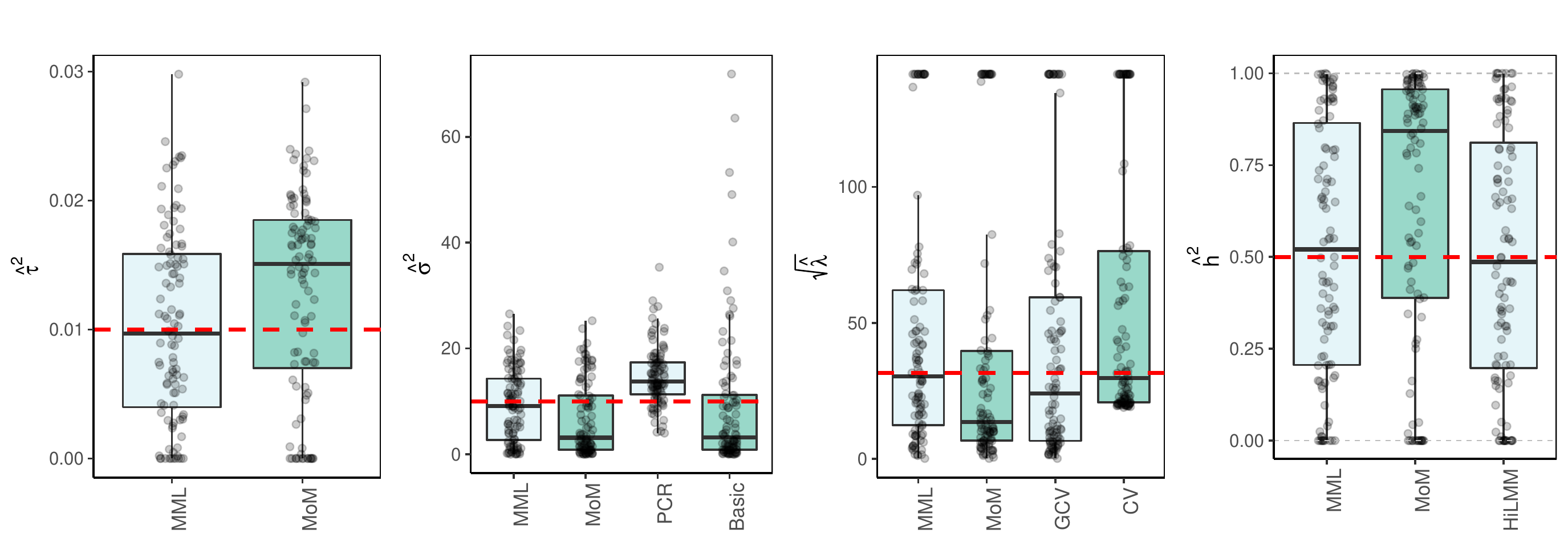}
  \caption{Gaussian $\beta$'s, $t_4$ errors.}
\end{subfigure}
  \label{fig:robust}
\end{figure}

\subsubsection{Binomial ridge regression}
\begin{figure}
\centering
\begin{subfigure}{.5\textwidth}
  \centering
 \caption{Binomial($N=5$) for independent $\BX$}
  \includegraphics[scale= 0.5]{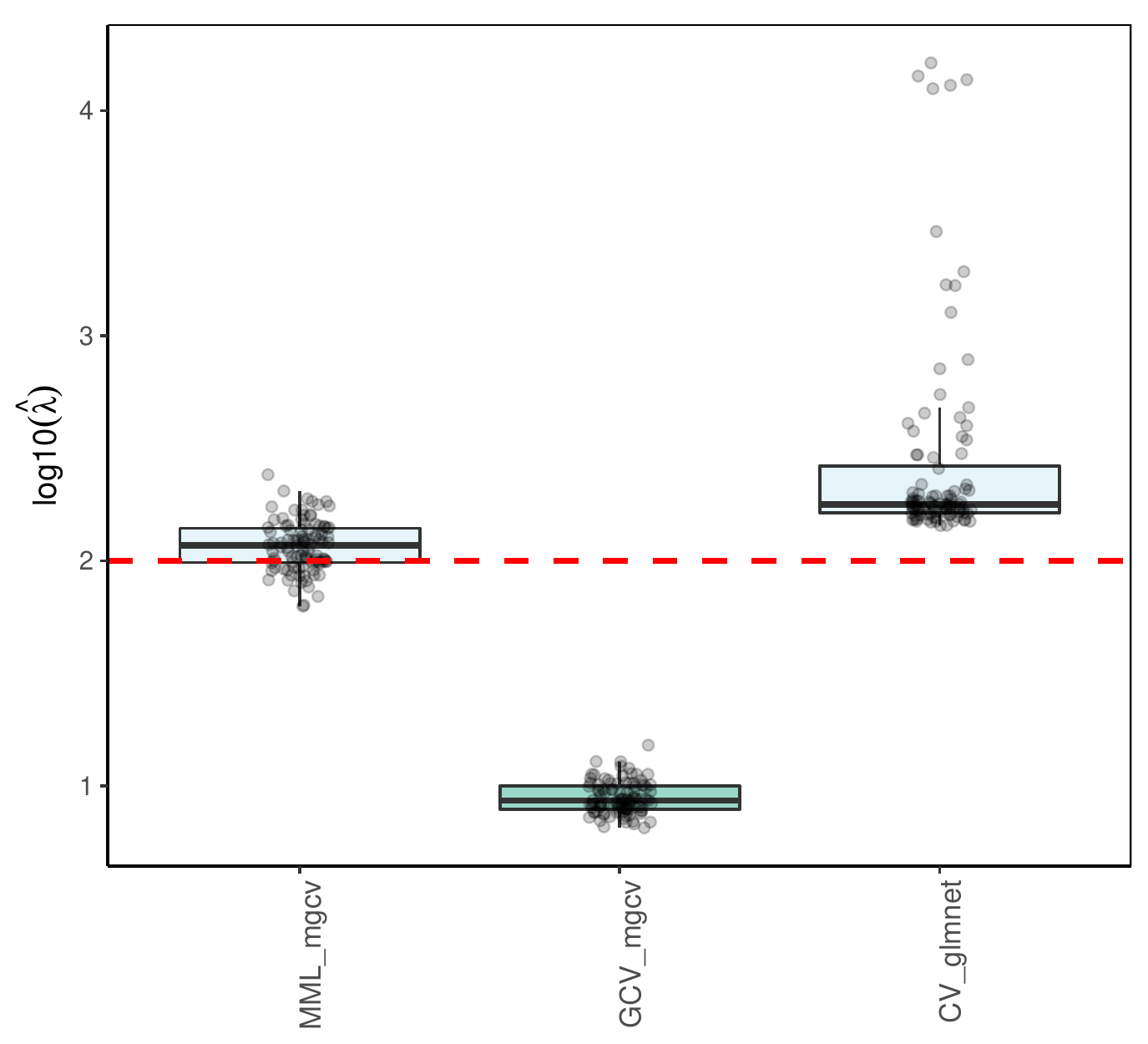}

\end{subfigure}%
\begin{subfigure}{.5\textwidth}
  \centering
   \caption{Binomial($N=5$) for multi-collinear $\BX$}
  \includegraphics[scale= 0.5]{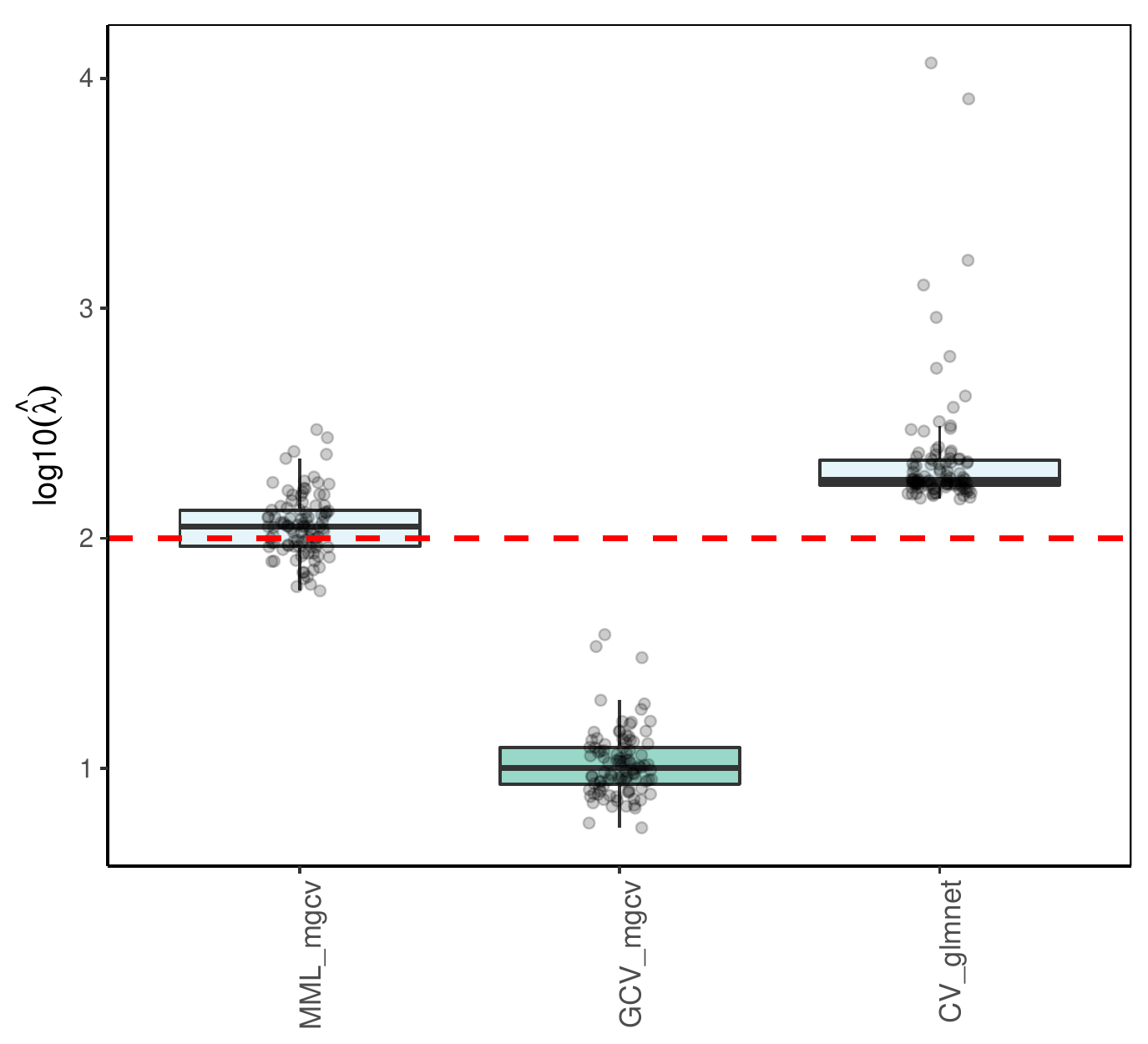}

\end{subfigure}
\caption{$\lambda$ estimates for Binomial($N=5$) ridge regression, $\lambda = 1/\tau^{2} = 100, n = 100, p = 1000$.}
 \label{Binomial}
\end{figure}

Note that unlike Poisson ridge regression Binomial ridge regression is not implemented in \texttt{penalized}. Hence, we compare \texttt{mgcv}-based results only with \texttt{glmnet}.
Figure \ref{Binomial} displays the results for estimating $\lambda$, in case binomial $N=5$. Results did not qualitatively differ for $N=3$ and $N=10$.
While MML is fairly good on target, the estimates from GCV are roughly 10 times too small, whereas the estimates from CV by \texttt{glmnet} are roughly 2-3 times too large,
with outliers towards a 10-100 fold overestimation.

\section*{References}

\bibliography{C://Synchr//Bibfiles//bibarrays}
\end{document}